\pdfoutput=1
\documentclass[preprint,journal]{vgtc}    

  \pdfoutput=1\relax                   
  \usepackage{graphicx}                
  \DeclareGraphicsExtensions{.pdf,.png,.jpg,.jpeg} 
  \usepackage{graphicx}                
  \DeclareGraphicsExtensions{.eps}

\graphicspath{{figures/}{pictures/}{images/}{./}} 

\PassOptionsToPackage{warn}{textcomp}  
\usepackage{textcomp}                  
\usepackage{mathptmx}                  
\usepackage[utf8]{inputenc}
\usepackage{mathtools, nccmath}
\usepackage{times} 
\usepackage{xpatch}
\xpatchcmd{\NCC@ignorepar}{%
\abovedisplayskip\abovedisplayshortskip}
{%
\abovedisplayskip\abovedisplayshortskip%
\belowdisplayskip\belowdisplayshortskip}
{}{}

\usepackage{cite}                      
\usepackage{tabu}                      
\usepackage{booktabs}                  
\usepackage{amsmath}

\usepackage{balance}  
\usepackage{graphics} 
\usepackage{times}    
\usepackage{url}      

\usepackage[breaklinks]{hyperref}

\usepackage[ruled,vlined]{algorithm2e}
\usepackage{enumitem}
\setlist{topsep=0pt, leftmargin=*}
\usepackage{xspace}
\usepackage[table,usenames,dvipsnames]{xcolor}
\usepackage{booktabs}
\usepackage{multirow}
\usepackage[10pt]{moresize} 
\usepackage{changepage}
\usepackage[mathscr]{euscript}
\usepackage{amssymb}
\usepackage{amsmath}
\usepackage{bm}

\usepackage{algorithmic, float}
\usepackage{blindtext}

\usepackage{stfloats}

\usepackage{MnSymbol}
\usepackage{microtype}                 
\microtypesetup{nopatch={footnote}} 

\newcommand{\arre}{A_{\mathrm{expectation}}}
\newcommand{\arrt}{A_{\mathrm{iter}}}
\newcommand{\arrv}{A_{\mathrm{value}}}

\onlineid{1289}

\vgtccategory{Application}

\newcommand{\etal}{{\it et~al.}\xspace}

\usepackage{comment}

\newcommand{\md}[1]{{\color{black}{#1}}}
\newcommand{\new}[1]{{\color{black}{#1}}}
\newcommand{\rv}[1]{{\color{black}{#1}}}

\newcommand{\name}{Calliope-Net\xspace}

\begin{document}

\title{Calliope-Net: Automatic Generation of Graph Data Facts via \\Annotated Node-link Diagrams}
\author{Qing Chen, Nan Chen, Wei Shuai, Guande Wu, Zhe Xu, Hanghang Tong, and Nan Cao}
 \authorfooter{
   \item
   	Qing Chen, Nan Chen, Wei Shuai, and Nan Cao are with Intelligent Big Data Visualization Lab at Tongji University. Nan Cao is the corresponding author.
    E-mails: \{qingchen,shuaiwei,nan.cao\}@tongji.edu.cn
   \item
   	Guande Wu is with New York University.
   	E-mail: guandewu@nyu.edu
   \item Zhe Xu and Hanghang Tong are with University of Illinois at Urbana-Champaign.
 }

\abstract{Graph or network data are widely studied in both data mining and visualization communities to review the relationship among different entities and groups. The data facts derived from graph visual analysis are important to help understand the social structures of complex data, especially for data journalism. However, it is challenging for data journalists to discover graph data facts and manually organize correlated facts around a meaningful topic due to the complexity of graph data and the difficulty to interpret graph narratives. Therefore, we present an automatic graph facts generation system, Calliope-Net, which consists of a fact discovery module, a fact organization module, and a visualization module. It creates annotated node-link diagrams with facts automatically discovered and organized from network data. A novel layout algorithm is designed to present meaningful and visually appealing annotated graphs. We evaluate the proposed system with two case studies and an in-lab user study. The results show that Calliope-Net can benefit users in discovering and understanding graph data facts with visually pleasing annotated visualizations.
}


\keywords{Graph Data, Application Motivated Visualization, Automatic Visualization, Narrative Visualization, Authoring Tools}

\pagenumbering{gobble} 

\maketitle

\thispagestyle{empty}


\section{Introduction}
Graph data are applied in a wide range of domains, from social network analysis~\cite{freeman2004development} to the study of disease transmission pathways~\cite{campbell2017detailed}. Over the past few decades, graph visualization has been extensively studied~\cite{herman2000graph,lee_task_2006}. In the field of data journalism, the growth of social networks and social media has drawn even more interest to the exploration of insightful patterns in graph data~\cite{nettleton2013data}. \new{Such patterns or insights can also be referred to as graph data facts that represent a particular type of numerical measure in a specific graph or subgraph.} Graph data facts could assist in understanding the relationship among different entities and groups. Moreover, it is important for journalists to effectively deliver and present graph data facts intuitively and expressively. However, existing tools built for graph visualizations may insufficiently support the whole fact discovery, organization, and presentation process. 

\md{For data journalists, deriving valuable graph data facts is not an easy task.} Mining graph data facts requires journalists to have adequate data analysis skills and domain knowledge. Even experienced analysts need to spend much time and effort in assessing the interestingness and significance of a huge number of potential graph insights. While prior work has studied how to automatically extract insights from multi-dimensional data~\cite{ding2019quickinsights}, graph data contain more complex information. To explore graph data facts, journalists usually need to utilize standard graph visualization software or develop specific systems and manually select interesting facts. Given the amount of graph data facts and the manual cost for such exploratory visual analysis, conducting this tedious task is time-consuming, especially when journalists have limited time and inadequate expertise or domain knowledge. 

\md{Once graph data facts have been explored, }
effectively and efficiently organizing and presenting such derived facts in graph visualizations are important to timely deliver core messages around a meaningful topic to the audience~\cite{lee2015more,srinivasan_augmenting_2019}.
Despite the extensive studies on graph visualization~\cite{nobre2019state}, support to help create visual narratives of graph data facts is limited. 
In addition, although graph visualization tools such as Gephi~\cite{bastian2009gephi} can convert graph data into visualizations, the exported results are unsuitable to be used in data-driven storytelling. 
Annotated visualization is a suitable narrative form given that graph data facts are highly related to the graph structure, which requires presenting the location information and semantic meaning simultaneously.
However, creating such annotated visualizations is a laborious task when the data becomes large and complex due to the huge design space for organizing visual elements and effective graph layouts. 

To tackle these two challenges, we present \name, a novel framework to create annotated visualizations with facts automatically discovered from graph data. The framework consists of a fact discovery module, a fact organization module, and a visualization module. The fact discovery module automatically extracts interesting patterns and formulates them into facts. The fact organization module gradually selects facts and organizes them around a meaningful topic, which is guaranteed by the topic-evidence-explanation structure. The visualization module presents the selected facts as a series of expressive annotations on the graph. An annotation-aware graph layout algorithm is designed to place the graph visualization and annotations properly.

With the proposed framework, we significantly reduce the efforts involved in discovering graph data facts to effectively create expressive presentations. We then developed a prototype system with interactive features such as editing to accommodate different user purposes. Finally, we conducted case studies to demonstrate the usability of our system and an in-lab user study to evaluate the quality of the annotated graphs generated by \name. 

\rv{The contributions of this paper are as follows:
\begin{itemize}
\itemsep -1mm
    \item We present a framework to create annotated node-link diagrams with data facts automatically discovered and organized from graph data.
    \item We develop an interactive prototype system, which is publicly available at \url{http://calliope-net.idvxlab.com}.
    \item We apply data with two case studies and a user study to demonstrate the usefulness of the system.
\end{itemize}
}

\section{Related Work}
\subsection{Data-driven Annotated Visualization}
Annotations play an important role in various genres of narrative visualizations~\cite{segel2010narrative} through guiding users' attention graphically and providing data context~\cite{kosara2013storytelling, choe2015characterizing, ren2017chartaccent}. Ren \etal~\cite{ren2017chartaccent} proposed a design space for annotated visualizations and identified the forms and the target types of annotations. To create annotations efficiently and effectively, various data-driven techniques and tools~\cite{bryan2016temporal,latif2021kori} have been developed. Contextifier~\cite{hullman2013contextifier} targeted visualizations in news articles and implemented a tool to automatically generate annotated stock line charts. 
\new{VIS Author Profiles~\cite{NLG_visualization} focused on using natural language text combined with visualizations to generate descriptions.} Bryan \etal~\cite{bryan2016temporal} studied the creation and placement of interesting annotations on the temporal layouts. Moreover, Brath \etal~\cite{brath2018automated} automatically generated annotations extracting the most relevant information. 
Kori \etal~\cite{latif2021kori} proposed a mixed-initiative interface that enabled the synthesis of text and charts through interactive references.

Although the above work enables data-driven annotation creation, the annotation generation for the graph data has been insufficiently touched. Bach \etal~\cite{bach2016telling} discussed how to use the visual expressiveness and familiarity of comics to communicate changes in dynamic graphs. As a result, DataToon~\cite{kim2019datatoon} was established as an interactive authoring tool that facilitates the creation of data comics with recommendations of four automatically-detected structural patterns.
\new{Latif \etal~\cite{latif2019EuroVis} introduced a system that could establish the linking between the textual and visual representations, offering an interaction-enriched experience.}
To alleviate the barriers to creating graph narratives, we provide a flexible generation and editing tool that automatically finds data facts both in attributes and structural patterns in the form of annotated visualizations.

\new{Despite annotations, several other techniques can also help understand graph data. Various graph layout algorithms are designed for aesthetics~\cite{purchase2000effective, bennett2007aesthetics} and can be applied in different scenarios~\cite{huang2005layout, pohl2009comparing}. Also, numerous tools~\cite{ellson2001graphviz,shannon2003cytoscape,bastian2009gephi,hansen2010analyzing} have been developed to facilitate graph visualization. For instance, Graphies~\cite{romat2019expressive} provided an interface for the smooth authoring of static communicative node-link diagrams. However, all these authoring tools only generate a single node-link diagram that shows the graph's overview. In this paper, we aim to tell interesting graph data facts around a meaningful topic from different granularities, including nodes, communities, and the graph as a whole.}

\subsection{Automatic Fact Generation and Organization}
The past few years have witnessed the development of automatic fact generation and organization in data mining and visualization communities~\cite{vartak_seedb_2015,demiralp2017foresight,ding2019quickinsights,lee2019avoiding}. For example, Foresight~\cite{demiralp2017foresight} used a rule-based visual analysis approach to extract and visualize insights from multi-dimensional data.
Ding \etal~\cite{ding2019quickinsights} presented QuickInsights, an automatic insight extraction system that discovers interesting patterns based on statistical measures including significance and impact. Demiralp et al. proposed a method that ~\cite{demiralp2017foresight} facilitates interactive exploration of large datasets through fast, approximate sketching. \rv{Law et al.~\cite{law2020characterizing} identified 12 types of automatic insights from 20 existing tools. 
To assess the automatic insights, Ding et al.\cite{ding2019quickinsights} introduced a unified formulation of insights and scoring metrics irrespective of the type. 
Since the number of insights or data facts grows exponentially with the number of data attributes, it is time-consuming to enumerate each potential fact. Thus, several new algorithms have been proposed to speed up the generation~\cite{luo2020ICDE,mafrur2018ACM}.}

Apart from generating meaningful data facts, recent work has studied how to organize data facts into fact sheets and various narrative visualizations~\cite{chen2023does}. For example, Datashot~\cite{wang2019datashot} organized the facts into related and meaningful topics.
Calliope~\cite{shi2020calliope} further investigated automatically generating visual stories based on a logic-oriented Monte Carlo tree search algorithm. 
However, previous work is limited to multi-dimensional data while our work focuses on the fact generation and organization of relational network data with topological features.

\section{Design Goals and Framework Overview}
To better understand the journalists' workflow and design choices in creating graph visualizations, we conducted in-depth interviews with four experienced data journalists, summarized the design goals from expert feedback, and presented the framework overview.

\subsection{Expert Interview}
\label{sec:ExpertInterview}
We conducted semi-structured interviews~\cite{lazar2017research} with four data journalists (E1-E4, all females). Each interview lasted approximately one hour and was conducted online via meeting software that enables screen sharing. All the interviewees had more than three years of experience in data journalism and expertise with graph data. We found them by first collecting graph-related data journalism on various news websites and then contacted the authors. We first introduced our research topic and the core concepts (i.e., graph data, data fact, annotation) to the experts. Next, we asked them a set of prepared questions under two topics: the general workflow and design choices for creating graph visualizations for data journalism. Example questions include, ``what is the general 
creation process?'', ``which tasks do you find difficult or tedious?'', and ``which visualization type would you prefer to use (i.e., node-link diagram, matrix) and why?''. Throughout the interview process, we asked follow-up questions if we noticed that an interviewee's answer is unclear or if we wanted to dig deeper into the details of their daily conduct. We summarize the interview results as follows.

 \textbf{Tools and workflows.} 
All the experts reported using various tools at different creation stages which involve graph data. The exploration and organization workflow relies largely on personal preferences and skills. 
However, they need to finish most of the data-driven exploration and creation by themselves in most cases due to the tight schedule and the cost to communicate and collaborate with co-workers. They sometimes leave the final design optimization to the specialists to improve the overall appearance. The most frequently used tools for data exploration and presentation are Gephi~\cite{bastian2009gephi} and Tableau~\cite{chabot2003tableau}. Two typical workflows are observed in their creation process. One is that they have several potential themes in mind and then validate or enrich their assumptions with evidence found in graph visualizations. The other is that they are given some interesting datasets and need to explore the data and organize a meaningful topic around the explored data facts.

 \textbf{Visualization styles and formats.} 
All the experts preferred node-link diagrams for visualizing graph data due to their natural representations of relationships between nodes, and flexibility for customization. 
E3 and E4 mentioned other graph representations and commented that ``\textit{compared to arc diagrams, circle layout, the force-directed layout of node-link diagram is more intuitive to show the distance between nodes.}'' 
They also used text annotations frequently after exploration to mark down important information and present more details. Providing text annotations is important since readers can obtain the information directly from the visualization. Titles and other supporting text are commonly added to facilitate understanding. 

 \textbf{Difficulties and Suggestions.} 
Creating graph-related data news presents several difficulties, such as making sense of unfamiliar datasets within a limited time range and the labor-intensive process of creating visualizations.
The experts often use Gephi to generate the node-link diagram and then import it to vector editing tools such as Adobe Illustrator to add annotations and modify the style.
E2 commented that ``\textit{those graph visualization tools are quite helpful, but they are not 'intelligent' enough. For example, some common functions such as node search are hidden deep in Gephi and require an additional workload to search.}'' 
All the interviewees agreed that a certain level of automation would facilitate and inspire their creation. 
E3 emphasized that ``\textit{insight suggestion would be useful but the premise is that the algorithm is reliable.}''
Although those journalists who create data-driven news articles are open-minded and willing to apply state-of-art technologies or advanced tools, they believe that flexible interactions are important in daily practice. 


\subsection {Design Goals}
Based on the above analysis, five design goals are proposed.

\noindent
\textbf{G1. Extract interesting graph data facts.} 
Graph data facts are the building blocks of annotated charts. Extracting interesting data facts is a challenging task due to the complexity of the data and the abstractness of the fact model. To ensure the quality of the result, it is necessary to make sure that the facts extracted are credible and interesting.

\noindent
\textbf{G2. Organize facts into meaningful topics.}
An appropriate topic is essential to attract readers' attention and increase their memorability~\cite{kosara2013storytelling}. Common narrative topics around networks usually derive from a set of key nodes or groups and their associations with each other~\cite{bounegru2017narrating}. It is important to organize graph facts into meaningful topics.

\noindent
\textbf{G3. Design expressive and intuitive annotations.} 
Intuitive and expressive annotations are useful to emphasize important components on a graph, distinguish different types of facts, and provide explanations and context for the facts. Too much or too complex annotation designs would bring extra cognitive load for general audiences. Therefore, it is important to design intuitive and expressive annotations to deliver the facts in an easy-to-understand manner.

\noindent
\textbf{G4. Present the annotated graph with good readability.}
Graph layout is crucial to improve readability. In addition to the aesthetic considerations such as edge crossing, visual information flow, and annotation overlaps should also be considered. Overlapping can ambiguity, while annotations need to be placed near the related facts. The layout should be adjusted accordingly when graph facts are added or deleted. Thus, an effective layout algorithm is required to improve readability.

\noindent
\textbf{G5. Enable flexible user interaction.} 
\rv{From the interviews, we found that a fully automated approach can barely meet all the user requirements. Previous work on human-machine collaborative systems also proved that flexible interactions are important to support further editing and adjustments based on user equirements~\cite{deodhar2022humanml, Automated_Infographic}.} 
The system should enable users to explore and modify the annotated content and styles to meet specific scenario requirements. Apart from manual adjustments, recommendation functions such as potentially insightful facts could also be provided to facilitate interaction.

\subsection{Framework Overview}
\begin{figure}[t!]
    \centering    \includegraphics[width=\linewidth]{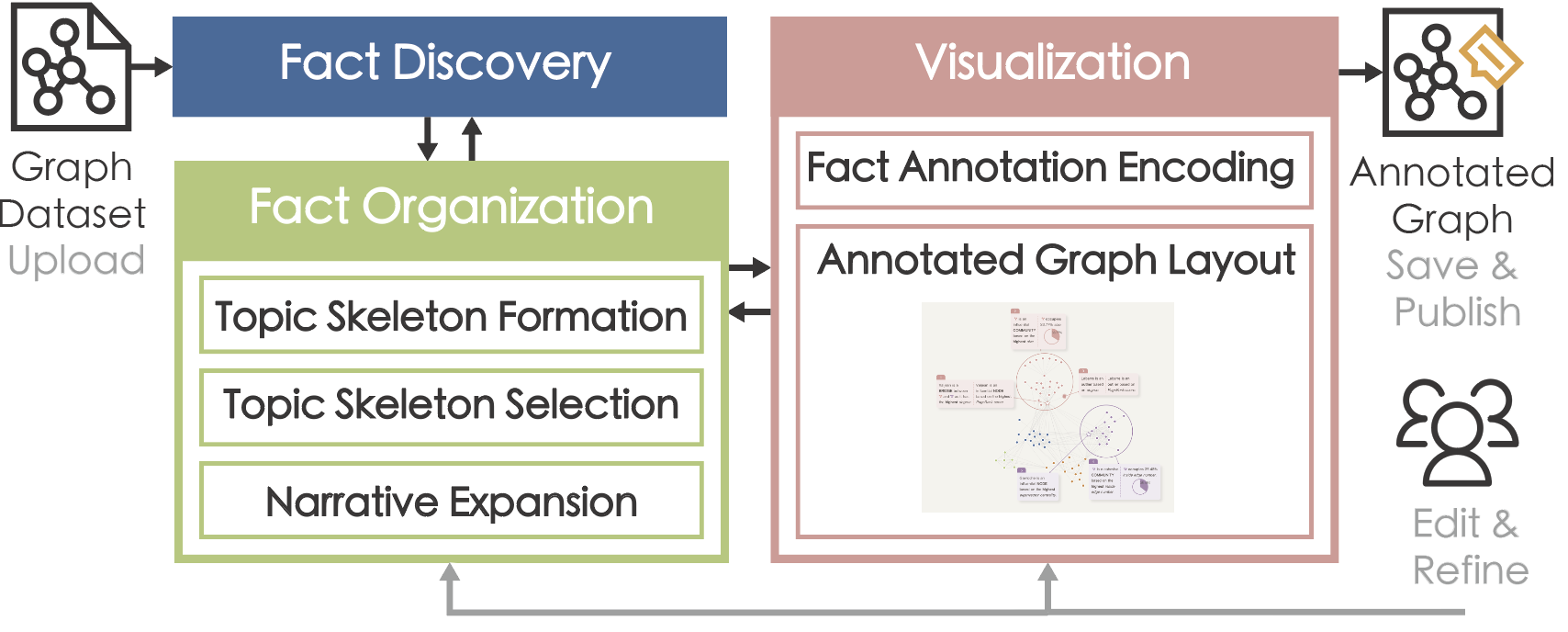}
    \vspace{-6mm}
    \caption{The architecture of the \name framework. }
    \label{fig:pipeline}
    \vspace{-5mm}
\end{figure}

\name is a web-based application comprising an automatic annotated graph generation engine and an editor. To generate an annotated graph, the user needs to upload graph data in JSON format, then view and refine the result automatically generated in the editor, and share and download (Fig.~\ref{fig:pipeline}).
The \name framework includes three core modules: a fact discovery module, a fact organization module, and a visualization module.
The fact discovery module extracts interesting patterns from original graph data and formalizes them into the facts (\textbf{G1}). 
The fact organization module gradually selects interesting facts and organizes them around a meaningful topic guaranteed by the topic-evidence-explanation structure \cite{kosara2013storytelling} (\textbf{G2}). 
After fact organization, the visualization module presents the data facts as expressive annotations (\textbf{G3}) and enables an annotation-aware graph layout algorithm to place the annotations in the graph properly (\textbf{G4}).
According to user preferences, the data facts, the fact order, text, graph and annotation layout, and style can all be edited and refined (\textbf{G5}).

\section{Fact discovery and organization}
\label{sec:factdiscovey}
\md{The fact discovery module is responsible for generating graph data facts that contain specific data patterns and form the cornerstone of our annotation generation. 
We initiate this section by introducing the concept of graph data facts, subsequently elaborating on our fact selection and organization methodology. 
}



\subsection{Graph Data Facts}
Graph data include a set of nodes connected by edges. Both nodes and edges can have several numerical or categorical attributes. Graph data facts can pertain to node-level facts, group-level facts (clusters), or graph-level facts. Here, we consider the node attributes and take a bibliography network as an example, where nodes represent the researchers with some attributes (e.g., id, country, citations) and edges denote the co-authorship between researchers.

\subsubsection{Fact Formulation}
Inspired by the concepts introduced in~\cite{wang2019datashot,shi2020calliope}, a graph data fact represents a particular type of numerical measure in a specific subgraph. The nodes on the subgraph can be partitioned into communities. Formally, a graph data fact $f_i$ is denoted by a 5-tuple:
\useshortskip \begin{equation*}
    f_i = \{type, subgraph, partition, measure, focus\}
\end{equation*}

\textbf{Type} describes the type of information involved in the graph analysis tasks~\cite{lee_task_2006,pretorius2014tasks,ahn2013task} and fact taxonomy~\cite{yang_chen_toward_2009}. 
We collected 47 explanatory node-link diagrams for graph data from data journalism and case studies of graph visualization papers, split them into 216 fact pieces, and calculated the distribution of different fact types in Table~\ref{tab:factTypes}. We then summarized the six common fact types (e.g., \textit{Extreme}, \textit{Outlier}, \textit{Rank}, \textit{Proportion}, \textit{Distribution}, \textit{Evenness}). Detailed descriptions of each fact type are provided in Table~\ref{tab:fact-overview}. 
As for type \textit{Value}, we found that it generally appears at the beginning of the collected cases to describe the number of nodes and edges, so we introduced the number of nodes and edges in the first sentence of the text summary.
As a start for graph data fact extraction, we only considered static graphs as our input, the temporal-related fact types (e.g., \textit{Temporal}, \textit{Trend}) are not included.
\begin{table}[H]
\centering
\caption{Distribution of different fact types}
\renewcommand\arraystretch{1.0}
\linespread{1.0}
\setlength\aboverulesep{0pt}
\setlength\belowrulesep{0pt}

\begin{tabular}{ccc|ccc}
\hline
\textbf{Fact Type} &  \textbf{Count} & \textbf{Ratio} & \textbf{Fact Type} & \textbf{Count} & \textbf{Ratio} \\ \hline
Extreme                             & 75                              & \multicolumn{1}{c|}{34.72\%}    & Trend                               & 13                              & 6.02\%                          \\ \hline
Value                               & 50                              & \multicolumn{1}{c|}{23.15\%}    & Proportion                          & 10                              & 4.63\%                          \\ \hline
Temporal                            & 21                              & \multicolumn{1}{c|}{9.72\%}     & Evenness                            & 9                               & 4.17\%                          \\ \hline
Outlier                             & 13                              & \multicolumn{1}{c|}{6.02\%}     & Rank                                & 8                               & 3.70\%                          \\ \hline
Distribution                        & 13                              & \multicolumn{1}{c|}{6.02\%}     & Difference                          & 4                               & 1.85\%                    \\ \hline     
\end{tabular}

\label{tab:factTypes}
\end{table}
\definecolor{mtcolor}{RGB}{223, 159, 157}
\definecolor{subspacecolor}{RGB}{107, 148, 104}
\definecolor{measurecolor}{RGB}{174, 133, 175}
\definecolor{dvcolor}{RGB}{214, 126, 58}
\definecolor{partitioncolor}{RGB}{182, 53, 43}
\definecolor{focuscolor}{RGB}{54, 108, 157}

\begin{table*}[tp]
\renewcommand\arraystretch{1.3}
\linespread{1.5}
\setlength\aboverulesep{0pt}
\setlength\belowrulesep{0pt}
\centering
\caption{Fact Overview. 
}
\begin{tabular}{|p{0.10 \textwidth}|p{0.26 \textwidth}|p{0.32 \textwidth}|p{0.20 \textwidth}|}
\toprule
\textbf{Type}  &Description 
& \textbf{Scoring Function}  &  Example        \\ \hline
Extreme    &Extreme facts refer to the corresponding community/node of the maximum feature value. &1. Extract the feature vector of all nodes/communities.
2. Calculate the interestingness score by the method used in \cite{shi2020calliope}. & {Node A is a super connector as it has the most connections.} \\ \hline Rank 
& Rank facts share a similar no- tion with the extreme while Rank facts focus on top-3 nodes/communities. &
1. Extract the feature vector of all nodes/communities.
2. Calculate the interestingness score by the method used in \cite{shi2020calliope}.  & {Communities A, B, and C are influential communities as they have top-3 sizes.}
 \\ \hline
Outlier    &Outlier facts measure how a node deviates from its neighborhood and the whole graph. 

&  
1. Calculate the neighborhood's average score $\tilde{x}_i$ of each node $i$.
\newline 2. Define the node $i$'s interestingness score as $\frac{|\tilde{x}_i-x_i|}{\max(\tilde{x}_i,x_i)}$.  & {Node A is an isolated community based on scarce external connections}.
\\ \hline
Proportion  &Proportion facts measure the extent to which the leading value dominates the feature entries.   
& 
1. Sort the feature vector in descending order to obtain the maximum value $x_{max}$.
\newline 2. Define the interestingness score as $\min{(1, \frac{2\times x_{\texttt{max}}}{\sum x})}$.  & {Community A occupies 18.8\% of the total size.}
 \\ \hline
Distribution  & Distribution facts are to determine if the feature vectors are well-modeled by a Gaussian distribution. 
& 
     1. Perform the Shapiro–Wilk test on the given feature vector to obtain the $p$-value.
    \newline 2. Define the interestingness as $1-p$.  & {The histogram shows the distribution of the communities based on the values of size.} 
 \\ \hline
Evenness    
&Evenness facts determine if the entries from the feature vector are uniformly distributed.  
& 
 1. Perform the Chi-squared test against the null hypothesis that the feature vector should be a constant vector, to obtain the $p$-value.
\newline 2. Define interestingness as $1-p$. & {The communities have an even distribution based on the values of size.}
  \\ \bottomrule
\end{tabular}
\label{tab:fact-overview}
\vspace{-6mm}
\end{table*}

\textbf{Subgraph} is defined as a set of data filters that zoom into a subgraph from the original graph in the following form: \useshortskip \begin{equation}
    \{\{\mathcal{F}[1]:\mathcal{V}[1]\},...\{\mathcal{F}[n]:\mathcal{V}[n]\}\}
\end{equation}
where $\{\mathcal{F}[1]:\mathcal{V}[1]\}$ denotes a filter item with particular value $\mathcal{V}[i]$ on a specific field $\mathcal{F}[i]$. 
The field $\mathcal{F}[i]$ can be the graph partition algorithm (i.e., \textit{Connected components} or \textit{Greedy modularity community}~\cite{clauset2004finding}) in addition to the categorical node attributes (\textbf{G1}), and filed $\mathcal{F}[i]$ can take the same value as filed $\mathcal{F}[j]$ as long as $\mathcal{V}[i] \neq \mathcal{V}[j]$. Since even if the components are selected by different values, they are still connected by edges. 
For example, {\it \{\{country=China\},\{country=Japan\}\}} is a subgraph that includes Chinese and Japanese researchers.

\textbf{Partition} is used to divide the subgraph into a set of components (i.e., nodes or communities) based on the categorical node attributes or the graph partition algorithms (\textbf{G1}). To analyze the graph data from different granularities, \name supports exploration from both node level and community level. By default, the partition is none (i.e., node-level), where we treat each node as an independent component. 

\textbf{Measure} is a numerical node attribute or a topological feature (e.g., node degree, community density) (\textbf{G1}). For the node-level topological feature, we calculate the values of degree, PageRank score\cite{flake2000efficient}, and eigenvector centrality\cite{bonacich1987power}. For the community level, we support size, inside edge number, density, average degree, average out-degree fraction, maximum out-degree fraction\cite{flake2000efficient}, separability~\cite{shi2000normalized} and cut ratio~\cite{fortunato2010community}.

\textbf{Focus} highlights one (or some) of the partitioned components in the subgraph. For instance, for the \textit{Outlier} fact type, the focus component is the node with an outlier value.

The graph data facts bear a resemblance to the facts extracted from multi-dimensional data~\cite{wang2019datashot,shi2020calliope} in terms of the 5-tuple construction, while presenting several major differences concerning the topology of the graph. 
First, \name can select the subgraph based on topology information (e.g., one of the components divided by connected components).
Second, \name can explore the graph from different granularities (e.g., node-level, community-level).
Third, \name can extract topological features (e.g., node degree, community density) from different components of the graph.
For example, the graph data fact, {\it \{Extreme, \{\{country=China\}, \{country=Japan\}\}, \{none\}, \{degree\}, \{id=Wang\}\}}, describes that \emph{Wang} is an influential researcher based on the \emph{highest degree} between \emph{China} and \emph{Japan}. 
Similarly, {\it \{Distribution, \{\{country=China\}\}, \{connected component\}, \{count\}, \}}, divides the researchers in \emph{China} into various \emph{connected components} and studies the \emph{distribution} of different researcher groups' \emph{size (count)}.

\subsubsection{Fact Scoring}
To extract meaningful and interesting facts (\textbf{G1}), we evaluate the importance of a graph data fact $f_i$ from two perspectives, \textbf{\textit{impact}} and \textbf{\textit{interestingness}}, based on the previous research~\cite{ding2019quickinsights,tang2017extracting} on auto-insight. \rv{The \textit{impact} of a fact is determined by the extent of its involvement with the nodes or edges in the related subgraph, relative to all the nodes or edges in the entire graph. In other words, the greater the number of associated nodes and edges, the higher the \textit{impact} value of the fact.} $G(V,E)$ denotes the input graph $G$ where $V$ is a set of nodes and $E$ is a set of edges. $G_{i}(V_{i},E_{i})$ represents the selected subgraph of $f_i$ where $V_{i} \subseteq V$ and $E_{i} \subseteq E$. 
Thus, \name obtains \textbf{\textit{impact}} as follows:
\useshortskip \begin{equation}
    impact(f_i)=\max(\frac{|V_{i}|}{|V|},\frac{|E_{i}|}{|E|})
\end{equation}
where $|\cdot|$ denotes the number of elements of the given set. 
For the \textbf{\textit{interestingness}}, \name evaluates it based on the statistical properties with a fact-type-specific function. We follow the idea of existing auto-insight work~\cite{ding2019quickinsights,tang2017extracting,wang2019datashot,shi2020calliope} to design the function $interestingness(f_i)$.
The detailed descriptions of the $interestingness(f_i)$ score for different fact types are provided in Table \ref{tab:fact-overview}. 
Since the facts that are both impactful and interesting would be considered meaningful, the final \textbf{\textit{score}} of graph data fact is computed as: 
\useshortskip \begin{equation}
    score(f_i) = impact(f_i)\times interestingness(f_i)
\label{eq:score}
\end{equation}

\subsubsection{Fact Annotations}
\rv{\name present factual content with text annotations to ensure easy understanding. Instead of using technical terms, we employ accessible descriptions. Through a thorough study of data-driven news, we utilized graph visualizations and relevant references to efficiently summarize common meanings of technical terms. We carefully selected common words or phrases to enhance accessibility.

For example, we describe (a) a node with the highest degree as ``a super connector'', (b) a node with the highest PageRank score as ``an important node because it has many connections and important neighbors'', (c) a node with the highest eigenvector centrality as ``an influential node while considering both direct and indirect connections'', (d) a community with a high inside edge number as ``having a close internal relationship'', (e) a community with a high density or average degree as ``highly connected'', (f) a community with high average-out degree fraction or maximum-out degree fraction as ``having strong external connectivity'', (g) a community with a high separability as ``an isolated community with scarce external connections''.} \md{The detailed explanations and examples are also provided on the system website.}

\subsection{Fact Selection and Organization}
\rv{The narratives around a static network can be classified into four topics, namely, \textit{exploring associations around single actors}, \textit{detecting key players}, \textit{mapping alliances and oppositions}, and \textit{revealing hidden ties}~\cite{bounegru2017narrating}.}
To organize facts into these four meaningful topics (\textbf{G2}), 
\name employs a topic-evidence-explanation structure proposed by 
Kosara and Mackinlay
\cite{kosara2013storytelling} to gradually select relevant facts and complement the narratives. \md{The narrative structure as shown in Fig.\ref{fig:algo}(2) starts with a topic claim (denoted as red nodes), followed by a series of evidence facts (denoted as green nodes), which are then elaborated upon by explanation facts (denoted as grey nodes).} This structure guarantees an inverted pyramid structure that guided the readers towards the most important information in the narrative, a technique commonly applied in data journalism~\cite{hart2021storycraft} and visual storytelling~\cite{knaflic2015storytelling}.

\subsubsection{Structure Overview}

\begin{figure}[t!]
    \centering
    \includegraphics[width=\linewidth]{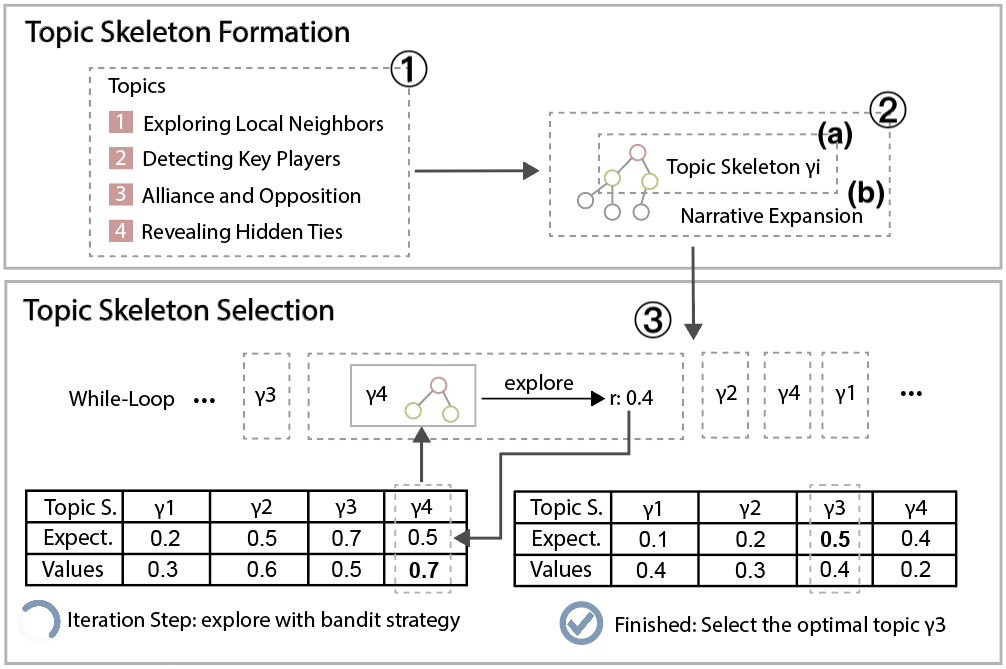}
    \vspace{-3mm}
    \caption{The formation and selection of the topic skeleton.}
    \label{fig:algo}
    \vspace{-6mm}
\end{figure}

\md{
To maintain a topic-evidence-explanation structure, we ensure that each graph narrative strictly aligns with a specific topic shown in Fig.\ref{fig:algo}(1)~\cite{bounegru2017narrating}. We introduce the concept of \textit{``Topic Skeleton''}, which comprises facts that are primarily related to the chosen topic. It is important to note that the topic skeleton can vary depending on the narrative topic. For instance, within a StackOverflow data network in Fig.\ref{fig:casestudy}(b), the abstract topic \textit{Exploring local neighbors} could be realized in topic skeletons such as ``Exploring C++'s local neighbors'' and ``Exploring Javascript's local neighbors''.
For the implementation of topic skeletons, \name employs a tree structure to arrange the facts, in which nodes signify graph data facts and links embody the sequential relations between these facts. Typically, the root fact (the red node in Fig.\ref{fig:algo}(a)), which is the fact most related to the topic, forms the foundation of this structure. Other fact nodes (the green nodes in Fig.\ref{fig:algo}(a)) then branch out from this root, serving as supportive evidence for this central fact. The specific mechanics of crafting this ``topic skeleton'' are expounded in Section~\ref{sec:graphtopics}. This arrangement shapes the topic-evidence part of the structure. To fully manifest the topic-evidence-explanation structure, further explanatory data facts (the grey nodes in Fig.\ref{fig:algo}(b)) can be incorporated into topic skeletons in the subsequent narrative expansion phase in Section~\ref{sec:expansion}, ultimately resulting in a complete graph narrative.
}
\subsubsection{Topic Skeleton Formation} 
\label{sec:graphtopics}
\md{
In graph narratives,}
\textbf{\textit{Relations}} among the extracted facts are based on both topological and logical considerations. The topological relation is specially designed for network narratives, where facts sharing the same focus and facts with connected focuses are considered. For example, many data-driven news articles explore neighboring nodes of a single node~\cite{2016election, proxy}. Transitions between the facts based on logical relations contribute to effective narrative delivery. Thus, we leverage logical relations to present cohesion between graph data facts, including \textit{Similarity}, \textit{Contrast}, \textit{Elaboration}, and \textit{Generalization}. These four types of logical relations are summarized from the transition taxonomy by Hullman \etal\cite{hullman2013deeper} where \textit{Temporal}, \textit{Spatial} and \textit{Casual} transitions are excluded because they are not popular in the static network.
\rv{To be specific, \textit{Similarity} refers to the relation between two logically parallel facts. This is identified when the \textit{measure} in the 5-tuple graph fact is modified while retaining the same focus components.
\textit{Contrast} is identified when two facts contradict with each other, which is indicated by the deviation of the focus components when the \textit{measure} changes.
\textit{Elaboration} occurs when the successor fact provides more details and expands upon a predecessor fact. \name allows three types of elaboration: (1) from a global network to a subgraph, (2) from community-level to node-level facts, and (3) from an \textit{Extreme} or \textit{Rank} fact to a \textit{Proportion} fact, since the latter describes the proportion of the largest data item, which is also the focus component of the former. 
\textit{Generalization} happens when the 
successor fact generalizes the information presented in the predecessor fact. \name enables three types of generalization between facts, which is the reverse sequence of \textit{Elaboration}.}
Therefore, we define a relation as:
\useshortskip \begin{equation}
 r := \{f_p, f_s, r_l, r_t \} 
\end{equation}
where $f_p$ is the predecessor fact and $f_s$ is successor fact, and $r_l$ and $r_t$ are the logical and topological relations respectively.

\md{
In the following subsection, we explore topic skeletons for four common topics using existing literature and collected samples. 
\textbf{\textit{Exploring Local Neighbors}} involves showcasing an influential node alongside its neighboring nodes (as exemplified in \cite{proxy}). 
Initialization selects an influential node based on topological features like degree or eigenvector centrality. Facts topologically connected to this node are integrated into the skeleton. 
}
\textbf{\textit{Detecting Key Players}} explores the key nodes in the network based on the topological features and node attributes (e.g., \cite{bushmoney, socialpark}). \name includes two different skeleton ways for this topic, namely, \textit{Within-Measure} and \textit{Within-Subgraph} key players. For \textit{Within-Measure} key players, the algorithm identifies the extreme facts that have a mutual measure. For \textit{Within-Subgraph} key players, the algorithm enumerates all the available measures and finds extreme facts with different measures in a mutual subgraph. The subgraph is produced by a set of subgraph filters (e.g., all the nodes having a categorical attribute).
\textbf{\textit{Alliance and Opposition}} analyzes the dense communications inside communities \cite{watts1998collective} and the absence of communications between them in the network (e.g., \cite{london-riots, ttip}). \name describes the communities divided by a topological partition algorithm, which reflects the intra-community alliance relationship and the inter-community opposition relationship. Apart from the partition algorithm, \name specifies the central communities measured by a topological feature through a \textit{Rank} fact. The facts about other interesting communities can still be added in the \textit{narrative expansion} stage.
\textbf{\textit{Revealing Hidden Ties}} identifies the weak ties between two communities whose nodes are strongly tied internally (e.g., \cite{chris}). The two communities are selected from the communities partitioned by the categorical attributes or the graph partition algorithms appointed by a parameter. \name identifies the tying node which delivers an \textit{Extreme} fact measured by a topological feature in the subgraph of two communities that connects them. Finally, \name builds the skeleton with the facts from the corresponding communities.

\subsubsection{Topic Skeleton Selection}
\label{sec:selection}
\md{The topic skeleton formation process can generate a diverse range of topic skeletons. Each topic skeleton can be expanded into a complete graph narrative by adding more elaborative facts to optimize the rewards defined in Sec.\ref{sec:expansion}. Prior to commencing narrative expansion, we need to identify and select the most promising topic skeleton. However, enumerating all potential topics to identify the optimal one would be impractically time-consuming.
Thus, \name implements a \textit{bandit strategy}, to identify the most promising topic from a set of topic skeleton lists, denoted as $\Gamma_s = \{\gamma_1, \gamma_2,...,\gamma_N\}$, where $\gamma$ refers to a topic skeleton and $N$ is the total of the skeletons. This bandit strategy performs a series of iterative evaluations, on the candidates in $\Gamma_s$. In each iteration, it selects a topic skeleton $\gamma'$ for evaluation as shown in Fig.\ref{fig:algo}(3). This evaluation is based on its current knowledge of each skeleton's quality, balancing the need to explore less-tested skeletons with the desire to exploit ones that have performed well in previous evaluations. For example, given two candidates with the same expected quality, the bandit strategy might deem the less-explored one as more promising, as there is greater uncertainty (and thus the greater potential for undiscovered interesting facts) regarding its true quality.

\name maintains three arrays in the algorithm: (1) the \textit{expectation array $\arre$}, which stores the mathematical expectations of each topic skeleton, (2) the \textit{iteration time count array $\arrt$}, which tracks the number of evaluation iterations of each topic skeleton, and (3) the \textit{Quality value array $\arrv$}, which reflects the remaining value of exploring each topic skeleton, based on its corresponding value in $\arre$, and $\arrt$.
$\arre$ values are initially set to zero. Subsequent updates to these values are driven by the objective function detailed in Section~\ref{sec:expansion}.}
The algorithm starts by exploring each candidate once and initializing $\arre$ and $\arrt$ arrays.
It then runs iteratively by selecting the most promising topic skeleton to explore next, based on its corresponding quality values $\arrv$. 
The $explore$ function is responsible for randomly sampling a graph story and computing its reward, which is obtained by summing up the scores in Equation \ref{eq:score}.
The algorithm continues to run until the test time upper bound is reached, at which point it outputs the most promising topic skeleton $\gamma^*$, based on the average expectations. 

\rv{In the selection process, the algorithm employs a randomness-based search to approximate the objective function result and calculate the result expectations of different topic skeletons. 
By assigning more attempts to the promising topic skeletons, the algorithm saves time by avoiding the exploration of less important ones, and increases the likelihood of selecting the optimal one.
Next skeleton is chosen based on expected results and attempt the number of attempts so far.
Topic skeletons with fewer attempts are prioritized for later exploration.}

\subsubsection{Narrative Expansion}
\label{sec:expansion}
\rv{In this stage, we aim to optimize a given topic skeleton by augmenting it with additional relevant facts. The algorithm executes an iterative process until the objective function, denoted as $R_N$, can no longer be improved by the addition of more facts.
To begin with, the algorithm initializes a candidate set that contains relationships between the facts in the topic skeleton and those with logical or topological connections to them. During each iteration, the algorithm selects the most promising candidate relation that yields the highest value and adds it to the topic tree. 
Subsequently, the candidate set is updated based on newly added fact relations. The algorithm terminates and generates the final topic tree when no further facts can improve the objective function value. The objective function $R_N$ is defined as the sum of the three scores, i.e., \textit{fact scores}, \textit{narrative diversity}, and \textit{network integrity}. }  

\textbf{\textit{Fact score}} reflects the meaningfulness of the selected facts, which is formulated as follows:
\useshortskip \begin{equation}
F(N) = \Sigma_{f_i \in F_N} score(f_i)
\end{equation}
where $F_N$ refers to the fact set of the narrative $N$ and $score(f_i)$ is calculated by Equation ~\ref{eq:score}. \md{The fact score reflects the statistical significance and is commonly used in data mining \cite{ding2019quickinsights}.}

\textbf{\textit{Narrative diversity}} estimates the diversity of fact types about a focus component. As fact scores are calculated based on type-specific functions, diversity reward can partially reduce the bias caused by the different fact types. 
\md{The diversity reward can augment the expressiveness and variety of the output, corroborated by research in narrative visualization~\cite{shi2020calliope, bounegru2017narrating}.}
\useshortskip \begin{equation}
D(N) = - \Sigma_{focus \in FC_N, type \in T} \text{ } p(focus,type) \times ln\frac{p(focus,type)}{p(focus)}
\end{equation}
where $FC_N$ refers to the focus component set of the narrative $N$ and $T$ includes all the fact types in Table \ref{tab:fact-overview}.
$p(focus, type)$ calculates the probability of the occurrence of the $focus$ component and type while $p(focus)$ calculates the focus component $focus$'s probability.

\md{The \textbf{\textit{network integrity}} metric evaluates the extent of the graph story's coverage against the provided input graph. This assessment aids in determining how comprehensively the story overlaps with the underlying data, a crucial aspect for ensuring its thoroughness \cite{segel2010narrative, shi2020calliope}. Moreover, in graph data mining, the influence of a pattern is a significant factor that can also be appraised using this metric \cite{herman2000graph}.}
We compute the reward as the percentage of the covered graph vertices of the story $N$'s focus components,
\begin{equation}
    I(N) = \frac{\cup_{focus \in FC_N} focus.vertices} {|V|}
\end{equation}
Given the three rewards, we can present the objective function as,
\begin{equation}
    R(N) = F(N) + D(N) + I(N)
\end{equation}

\section{Visualization and System Design}
In this section, we describe how the visual elements are designed, placed, and the interactions applied in the prototyping system.


\subsection{Fact Annotation Encoding} \label{sec:visual}

Referring to the design space from \cite{ren2017chartaccent}, \textit{annotation target} and \textit{annotation form} need to be considered when designing annotations.   
\textit{Annotation target} is a data item/set (in our case, node/community), which is the focus of the data facts. Regarding the design of \textit{annotation form}, we use a combination of various annotation forms (i.e., text, highlights, glyphs as shapes, and visualizations as images) to emphasize the essential elements, distinguish different fact types and provide expressive explanations (\textbf{G3}).
As mentioned in Section \ref{sec:factdiscovey}, we summarized six fact types, four of which are with focus (e.g., \textit{Extreme}, \textit{Outlier}, \textit{Rank}, \textit{Proportion}) and two that are not (e.g., \textit{Distribution}, \textit{Evenness}). For a fact with focus, we highlight its focus (e.g., node, community) on the graph and connect the focus to the annotation with leader lines. When the focus is a node, we replace the node with a well-designed glyph to denote its fact type. \textit{Rank} and \textit{Proportion} facts are denoted by annotations with both visualizations and generated text descriptions, while \textit{Extreme} and \textit{Outlier} facts are presented with text-only annotations. For a fact without focus, we consider it as graph-level information and therefore present it as a captioned chart in the \textit{graph summary}.




\begin{figure}[t!]
    \centering
    \vspace{-1mm}
    \includegraphics[width=\linewidth]{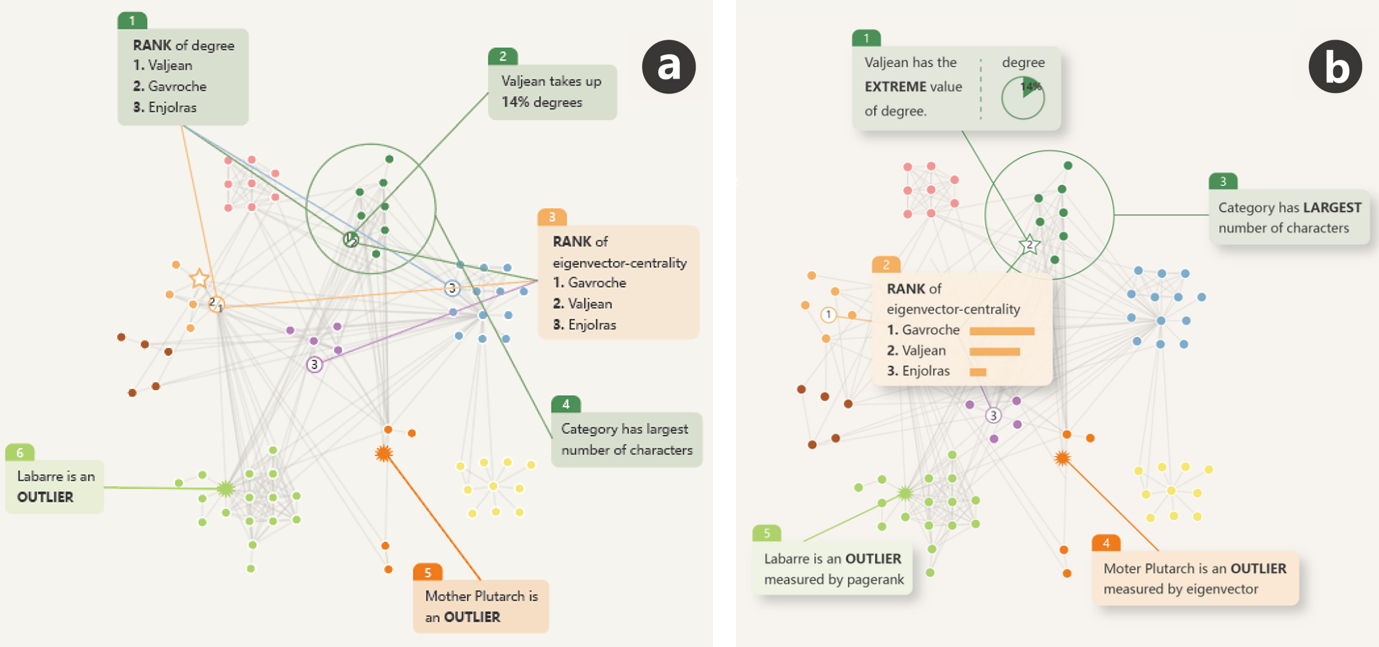}
    \vspace{-3mm}
    \caption{
    \new {Two outputs generated by \name: the layout before 
 the final optimization (a) and the layout after the final optimization (b) .} 
    }
    \label{fig:layout_optimization}
    \vspace{-5mm}
\end{figure}

\subsection{Annotated Graph Layout}

After presenting facts as annotations, we can get annotations with exact widths and heights. Therefore, the input for the layout algorithm consists of an undirected graph $G(V,E)$ where $V$ is a set of $v$ nodes and $E$ is a set of $e$ edges, a set of $m$ annotations $A={\{a_1, ..., a_m\}}$ of width $w_i$ and length $h_i$, and $n$ annotation leaders $L$ ($n\geq m$).
When the annotation describes a rank-type fact, three leader lines connect the annotation with the top three nodes or communities. 
We consider a node to be a \textbf{\textit{data element}} if an annotation depicts the node or the community it contains. Each annotation $a_i$ can correspond to one or multiple data elements $a_i^d$. 
To achieve the readability goal (\textbf{G4}), it is crucial to aesthetically place the annotations while visualizing the graph. Therefore, we design a layout algorithm that calculates the position $X_i$ for each node $i \in \{1,\dots, v\}$ and the center point $P_i$ for each annotation $i \in \{1,\dots, m\}$ to better illustrate the graph data. Our algorithm is carried out in three steps: (1) creating an initial graph layout according to visual information flows, (2) placing annotations in blank areas, and (3) optimizing the final layout to avoid overlap.

\new{Fig.~\ref{fig:layout_optimization}(a) shows the initial annotation layout according to visual information flows (circle) while Fig.~\ref{fig:layout_optimization}(b) shows the layout after the optimization. Compared to (a), (b) has minimized the leader line crossings, which helps relieve the visual burdens. The lengths of the leader lines in (b) are much shorter than those in (a), making it easier for users to capture the spatial relationship between the annotations and the focus nodes. The rank annotation lies in the middle of its three corresponding focus nodes (Gacroche, Valjean, Enjolras), enabling users to understand the information more efficiently.}

\subsubsection{Graph Layout} \label{sec:graphlayout}
To fulfill our design goal (\textbf{G4}), two constraints below are considered to improve the readability of the graph.

\textit{\textbf{Separated Communities.}} Communities are always prominent structures that draw users' interest, mainly when the annotated graph contains data facts of communities. Thus, we separate different communities in the graph layout to enhance the readability of their structures. 

\textit{\textbf{Visual Information Flow.}} To facilitate users to follow the order of the data facts, we adopt the concept of Visual Information Flow (VIF)~\cite{lu2020exploring} to guide the layout of the annotated graph. 
Considering diverse designs and arrangements,
the four design patterns, \textit{landscape}, \textit{clock}, \textit{left-wing} and \textit{right-wing}, were applied. Referring to~\cite{lu2020exploring}, each data element is expected to appear as an \textbf{\textit{anchor point}}.

Given a graph $G(V,E)$, annotations $A$, and leaders $L$, we need to select the most suitable VIF patterns.
First, we draw a graph by stress majorization~\cite{gansner2004graph} with high tolerance $\epsilon\sim10^{-4}$ to quickly achieve a rough layout, providing a general overview of the topology. 
Since we expect the data elements' positions to be placed as close as possible to the anchor point position, we 
use the center of the data elements to estimate the anchor point position.
Then, we match the connections of anchor points on the graph with each generated 
data fact
, based on four VIF patterns using Procrustes analysis~\cite{gower1975generalized} to satisfy translation, scaling, and rotation invariance and Fréchet distance~\cite{eiter1994computing} to measure the similarity. After selecting the VIF pattern, we can easily calculate the exact position of each anchor point. For example, when the VIF pattern is the \textit{clock}, the anchor points are evenly distributed on a circle with the graph's longest path as its diameter. 

Based on the constraints introduced above (separated communities, visual information flow), we solve our graph layout problem by minimizing the following stress function $S(\mathbf{X})$:

\useshortskip \begin{equation}
\label{eqn: stressfunction}
\begin{split}
S(\mathbf{X}) = &\sum_{i<j} w_{ij}||X_i - X_j - D_{ij}||^2 +\sum_{(i,j)\in E^{'}} w_{ij}^{'}||X_i - X_j - D_{ij}^{'}||^2 \\
&+ \sum_{i=1}^m\sum_{j \in a_{i}^d} w_{ij}^{a} ||X_j-A_i||^2 
\end{split}
\end{equation}
where $\mathbf{X}=\{X_i,...,X_v\}^\mathrm{T}$ is an $n \times 2$ matrix for all nodes.
The first term is the reformulated stress function~\cite{wang2017revisiting} of Stress Majorization~\cite{gansner2004graph}, where
$D_{ij}$ represents an edge vector whose length is the graph-theoretical distance between node $i$ and $j$ and $w_{ij}=D_{ij}^{-2}$ is the normalization constant that prioritizes nodes with small distance.
The second term aims to separate communities, where
$E^{'} = \{(i,j) | (i,j) \in E \text{, node } i \text{ and } j \text{ belong to two overlapping communities respectively}\}$,
$D_{ij}^{'}=(X_i - X_j) + MPD + \varepsilon$. Referring to~\cite{wang2017revisiting}, we detect overlap and then add the minimal penetration depth $MPD$ and constant separation parameter $\varepsilon$ to the corresponding target edge vector between them.
$w_{ij}^{'}$ (default as 0.5) is normalization weights.
The last term enables the nodes on the graph to be positioned as close as possible to the anchor points.
$A_i$ represents the position of the anchor point and the default values of the weights $w_{ij}^{a}$ is 0.5.
\subsubsection{Annotation Placement}
After computing the nodes' positions in the graph, we search for blank areas to place the annotations. Finding ``good" positions for the annotations is not an easy task because we need to calculate the two-dimensional coordinates of multiple annotations simultaneously. Thus, an efficient search algorithm named Harmony search~\cite{geem2001new} is adopted. The algorithm is a heuristic global search algorithm that has been successfully applied to a variety of combinatorial optimization problems. By this means, the problem of placing annotations is transformed into maximizing the quality function.

\textbf{Quality function.} 
\md{Inspired by previous work~\cite{kaneider2013automatic,wu2016focus+,bryan2016temporal} and based on our design goal (\textbf{G4}), we propose a function that evaluates the quality $Q$ of annotation placement by five metrics, node overlap (${m_1}$), disjoint annotations (${m_2}$), leader line crossing (${m_3}$), leader line length (${m_4}$), and visual information flow (${m_5}$)}:
\useshortskip \begin{equation}
Q = \sum_{i=1}^5 w_i \times m_i
\end{equation}
where $w_i \in [0,1]$, $\sum_i w_i = 1$ 
\md{are the weight parameters that we set empirically to balance different criteria. The default values of these weights are set to $\frac{3}{26}$, $\frac{3}{26}$, $\frac{10}{26}$, $\frac{5}{26}$, and $\frac{5}{26}$ respectively, based on our experiments.}
All the metrics are normalized to [0,1], where 0 indicates the worst quality, and 1 means the best.

\textbf{Minimize Node Overlap.} 
The node-link graph serves as the foundation, so annotations should not occlude nodes on the graph, especially those that are more significant. 
We use the degree centrality ${d_i}$ to measure the importance of the $i$-th node, $d_{max}$ denotes the maximum degree centrality. Other centrality metrics can also be used. 
The data elements are the most important nodes because they are the focus of the annotated visualization.
$T(i)$ denotes the $i$-th node's importance:
\useshortskip \begin{equation}
T(i) = \begin{cases}
    1& \text{if node is a data element}\\
    \frac{d_i}{d_{max}}& \text{otherwise}
\end{cases}
\end{equation}
Therefore, we look at the faction of the non-overlapping nodes, calculated as follows:
\useshortskip \begin{equation}
m_1 =  \frac{\sum_{i\in non\_overlap}T(i)}{\sum_{i=1}^{v}T(i)}
\end{equation}

\textbf{Disjoint Annotations.} To provide a smooth reading experience, the intersection area between the annotations should be as small as possible. $m_2$ is defined as the total area of displayed annotations divided by the sum of the areas of each annotation. The metric is 1 when no annotations overlap, which is the best-case scenario. 
 
\textbf{Minimize Leader Line Crossing.} 
Leader line crossings may cause visual disorder. However, crossing cannot always be avoided. When two leader lines are crossed, maximizing the angle between them can make them more identifiable.
Likewise, this value is normalized:
\useshortskip \begin{equation}
m_3 =  \frac{\sum C(l_{ij},l_{lk})}{n\times(n-1)/2},
\end{equation}
where $C(l_{ij},l_{lk})$ evaluates non-crossing score:
\useshortskip \begin{equation}
C(l_{ij},l_{lk})=\begin{cases}
    1& \text{lines $e_{ij}$ and $e_{lk}$ don't cross}\\
    sin\alpha& \text{$\alpha$ is the angle between lines $e_{ij}$ and $e_{lk}$}
    \end{cases}
\end{equation}

\textbf{Minimize Leader Line Length.} 
When the length of the leader line is too long, it is not easy to capture the spatial relationship between the annotation and the focus nodes. Therefore, the distance from the annotation (\textit{leader line target}) to its corresponding focus node (\textit{leader line source}) should be as short as possible.
$F_i$ donates the center of focus nodes. $s$ denotes the radius of the search range.
Therefore, $m_4$ can be measured using the following formula:
\useshortskip \begin{equation}
m_4 = 1 - \frac{\sum_{i=1}^{n}||P_i-F_i||}{n \times s}
\end{equation}

\textbf{Visual Information Flow.} Similarly, the distance from the annotation to the anchor point should not be too far, so that annotations can be well placed in the order of information flow. Thus $m_5$ is defined as:

\useshortskip \begin{equation}
m_5 = 1 - \frac{\sum_{i=1}^{m}||P_i-A_i||}{m \times s}
\end{equation}

\subsubsection{Position Optimization}
Although we considered the overlap of annotations ($m_1$, $m_2$) in the second step, it cannot be completely solved as the blank area is limited and the annotation size is much larger than the node. In the third step, we further optimize the layout to improve readability and aesthetics, considering removing overlaps between annotation and annotation, annotation and community, and annotation and node.
Similarly to Section~\ref{sec:graphlayout}, we remove overlap by minimizing the stress function. The first term in the equation (\ref{eqn: stressfunction}) is the same, while the second term is modified corresponding to different overlapping objects.

\begin{figure}[t!]
    \centering
    \vspace{-1mm}
    \includegraphics[width=\linewidth]{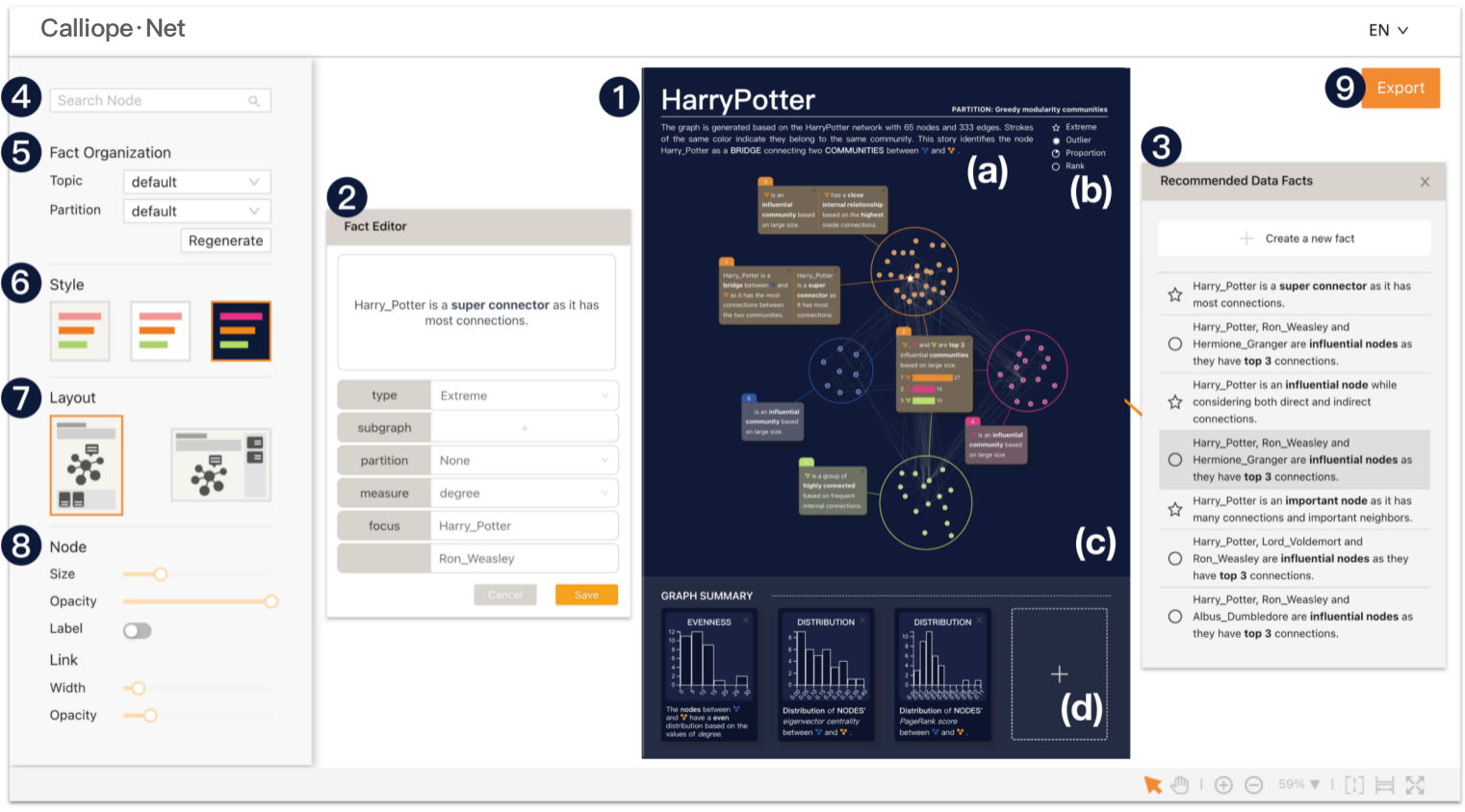}
    \vspace{-3mm}
    \caption{\rv{The editing page of \name system interface.
    }
    }
    \label{fig:interface}
    \vspace{-2em}
\end{figure}

\subsection{System Interface and Interaction}
We introduce the system interface and how to edit and refine the generated annotated chart (\textbf{G5}). To begin with, the user can upload the JSON format graph data and click the generation button. Then the main interface, consisting of the annotated graph and the configuration panel, is shown. 
In the annotated graph, the title and the text summary (Fig.~\ref{fig:interface}-1(a)) are shown at the top with the annotation legend (Fig.~\ref{fig:interface}-1(b)). 
The graph (Fig.~\ref{fig:interface}-1(c)) with annotations is shown at the center. \new{Facts in the annotations are captioned with generated text descriptions, and some facts are visualized with a chart.
Multiple facts can be presented in one annotation when they are related to the same node or community. }
The graph summary (Fig.~\ref{fig:interface}-1(d)), including distribution and evenness, is shown at the bottom. 
Users can interactively remove, edit and add a fact in the fact editor 
(Fig.~\ref{fig:interface}-2). \rv{By double-clicking on a particular node or community, users can check the recommended data facts focused on the node or community (Fig.~\ref{fig:interface}-3). The narrative order and text descriptions of the annotations can also be modified. In addition, users can manually adjust the locations of nodes, communities, and annotations.
To find a specific node, users can search the node name in the search box on the configuration panel (Fig.~\ref{fig:interface}-4).
In the fact organization zone (Fig.~\ref{fig:interface}-5), users can set the topics and the partition algorithm, and regenerate the annotated graph. Users can also change different styles and presentation layouts (Fig.~\ref{fig:interface}-6,7).
To further customize the annotated graph, users can modify the node and link attributes in detail(Fig.~\ref{fig:interface}-8). After all the adjustments are made, users can export the generated visualization by clicking the ``Export'' button (Fig.~\ref{fig:interface}-9).}

\section{Case study}

\begin{figure*}[t!]
    \centering
    \vspace{-2mm}
    \includegraphics[width=0.92\linewidth]{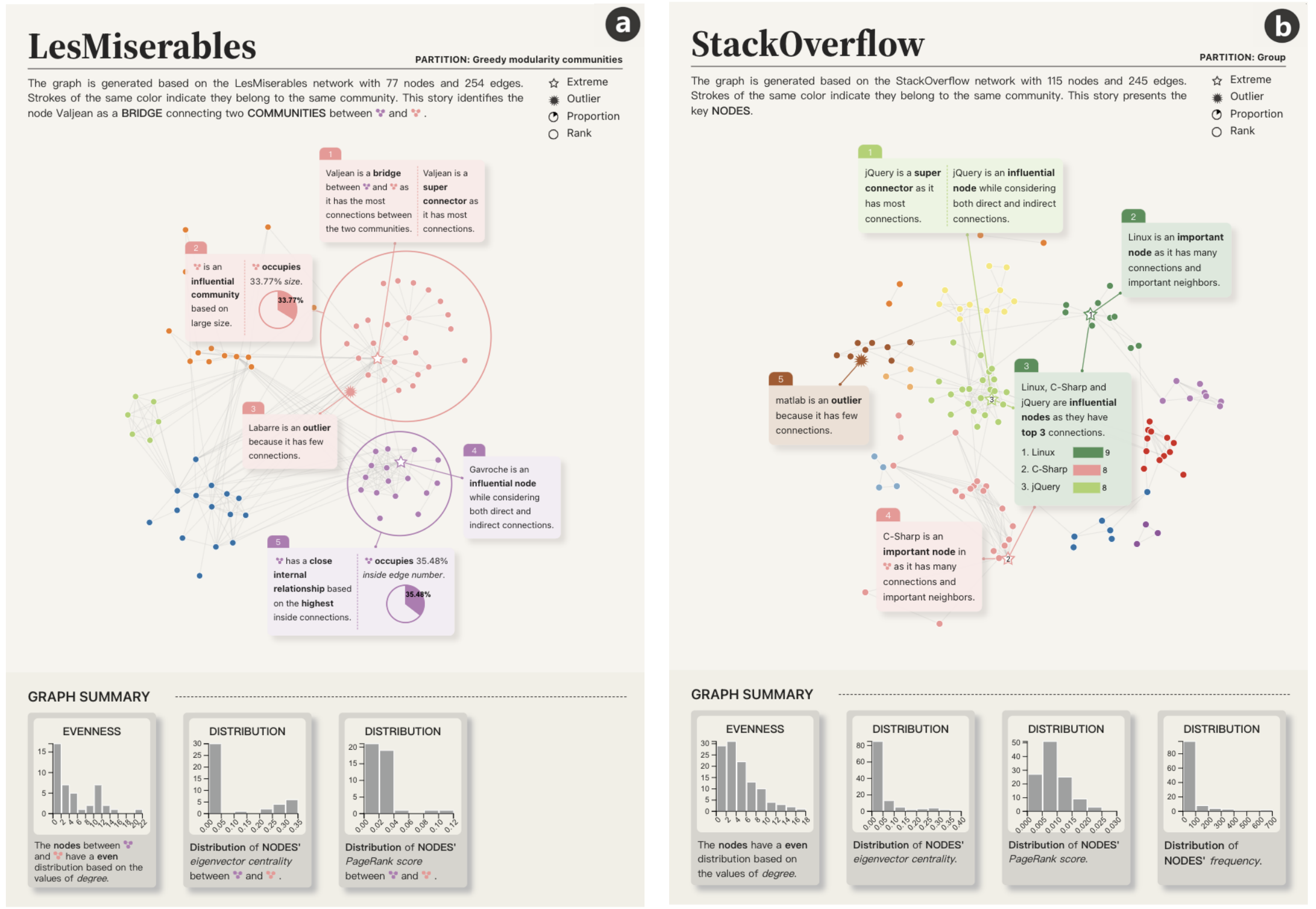}
    \caption{Two annotated graphs generated by \name in the case study: (a) presents the Les Misérables character relationship, and (b) shows the technology ecosystem by Stack Overflow Tag network.}
    \label{fig:casestudy}
    \vspace{-2mm}
\end{figure*}

\rv{To demonstrate the usefulness of \name, we performed case studies with two datasets. After an introduction and a quick demonstration of \name, we invited two experts, one visualization researcher with over 6 years of expertise in graph analysis (E5), and one data journalist with 5 years of experience in the field (E6), to explore the system and edit the generated graphs. The experts were encouraged to think aloud and ask any questions whenever they wanted. 

\subsection{Case Study One: Les Misérables}
The \textbf{\textit{Les Misérables Co-occurrence}} network dataset\cite{knuth1993stanford} has 77 nodes and 254 edges. Each node represents a character, and an edge connects two characters whenever they appear in the same chapter of Hugo's novel Les Misérables\cite{enwiki:1006098025}. 
With \name, the experts generated annotated graphs on two different topics,  \textbf{\textit{Exploring Local Neighbors}} and \textbf{\textit{Revealing Hidden Ties}}, respectively.
Fig.\ref{fig:casestudy}(a) shows a generated graph whose topic is \textbf{\textit{Revealing Hidden Ties}}. For example, the annotated graph illustrates that the protagonist, Valjean, is the bridge connecting two communities (in pink and purple) partitioned by Clauset-Newman-Moore greedy modularity maximization~\cite{clauset2004finding}, as he is the node with the most connections between two communities (\textit{Annotation 1}). 
The community containing Valjean is also quite influential, as it is the largest community with $33.77\%$ of the number of nodes in the entire network (\textit{Annotation 2}). 
Meanwhile, the purple community is a cohesive community based on the inside connections that accounted for $35.48\%$ (\textit{Annotation 5}). }
E5 agreed \new{that} the generated graphs contained interesting data facts, and the information annotated is convincing. 
Regarding the system, E5 commented that automatic annotations helped gain an initial sense of the network, 
\textit{``it's much smarter than I have imagined"}. 
The expert praised the easy-to-use generation and editing features. 
He was delighted to see those communities can be highlighted when hovering on and that communities can be moved and remapped. 


\subsection{Case Study Two: Stack Overflow}
\rv{The \textbf{\textit{Stack Overflow Tag}} network dataset \cite{stackoverflow} includes 115 nodes and 245 edges, where each node is a technology tag and an edge represents two tags appearing together on developers' \textit{Developer Stories}. With this dataset, the experts explored two annotated graphs generated by \name. }Fig.\ref{fig:casestudy}(b) shows one graph about \textbf{\textit{Detecting Key Players}}, aiming at discovering influential technologies in the tech ecosystem. 
\md{
Therefore, the generated graph highlighted the node jQuery with \textit{Annotation 1} as jQuery is a super connector with the most connections and it is also an influential node considering both direct and indirect connections. Linux is another important node as it has many connections and important neighbors (\textit{Annotation 2}). Then, the story continued on detecting other key players in the graph and find the three most influential nodes, i.e., Linux, C-Sharp, and jQuery based on the number of connections (\textit{Annotation 3}).
}
E6 enjoyed reading the generated graphs, \textit{``the generated graphs are quite cohesive, I can clearly see the insights are related and under the same topics''}. He also felt comfortable with the number of data facts included in the annotated graph, commenting that five to ten data facts displayed in around five annotations were appropriate.
In terms of visualization, E6 was satisfied with the graph design and spoke highly of the annotation and poster style. 
E6 observed that \textit{``it's interesting to identify the order in the ranking on the nodes''}. 
Regarding the prototype system, the expert acknowledged that the interactions are user-friendly for both generating and editing annotated graphs.
In the editing phase, he double-clicked to obtain a list of recommended data facts, \textit{``the fact recommendation mechanism is really helpful; otherwise, I would have spent more time searching for relevant data facts on my own}''.

\section{Evaluation}
We conducted an in-lab user study to evaluate the quality of the annotated graph visualizations generated by \name. 
Considering that \name is the first system that automatically generates annotated graphs directly from graph data, there is no baseline for comparison. Therefore, we compared the annotated graphs respectively generated by \name and created by domain experts based on the same datasets. 

\subsection{Data}
We collected four graph datasets from the Internet, including Harry Potter Co-occurrence network \md{(65 nodes, 333 edges)}, Game of Thrones Co-occurrence network \md{(96 nodes, 547 edges)}, Global Trade network \md{(144 nodes, 311 edges)}, and Political Books network \md{(105 nodes, 441 edges)}. 
For each dataset, we generated two annotated graphs by \name and invited the experts from previous interviews (E1-E4) to help create two distinct graphs. 
The experts used an editor that includes all the editing features on \name, but without any automatic functions such as fact generation and automatic annotation layout. 
After a 15-minute introduction, the experts were asked to explore interesting facts, add annotations, and manually adjust the final layout. Each annotated graph took about 25 minutes to create. 

\subsection{Procedure}
\md{
The recruitment process primarily utilized social media platforms as a promotional channel. The author specifically engaged with university students and student social media groups, considering the need for diverse backgrounds, to effectively distribute the advertisement.}
\rv{We recruited 36 participants (21 females) aged from 19 to 30 years old ($M=23.75$, $SD=2.42$). 13 participants are with over one year of experience in graph visualization (denoted as V1-V13), while the other 17 participants are without such backgrounds (denoted as P1-P17). }
Each participant was presented with four annotated graphs from four different datasets, two by \name and two domain experts. 
To achieve a fair comparison, the order of the datasets is fixed, but the presentation order of the annotated graphs created by the expert or \name is counterbalanced. 
For each annotated graph, the participant was asked to read carefully and rate the quality of the annotated graph on a 5-point Likert scale from the following aspects: (1) the data fact quality (\textit{Insightful, Comprehensive}), (2) the placement of the annotations (\textit{Reasonable}) and the annotated graph's layout (\textit{Aesthetics}), (3) the overall reader experience (\textit{Comprehension, Engagement, Memorability}). The participants were asked to provide comments on their ratings. Each participant took around 60 minutes to finish the study.

\begin{figure}[t!]
    \centering
    \includegraphics[width=\linewidth]
    {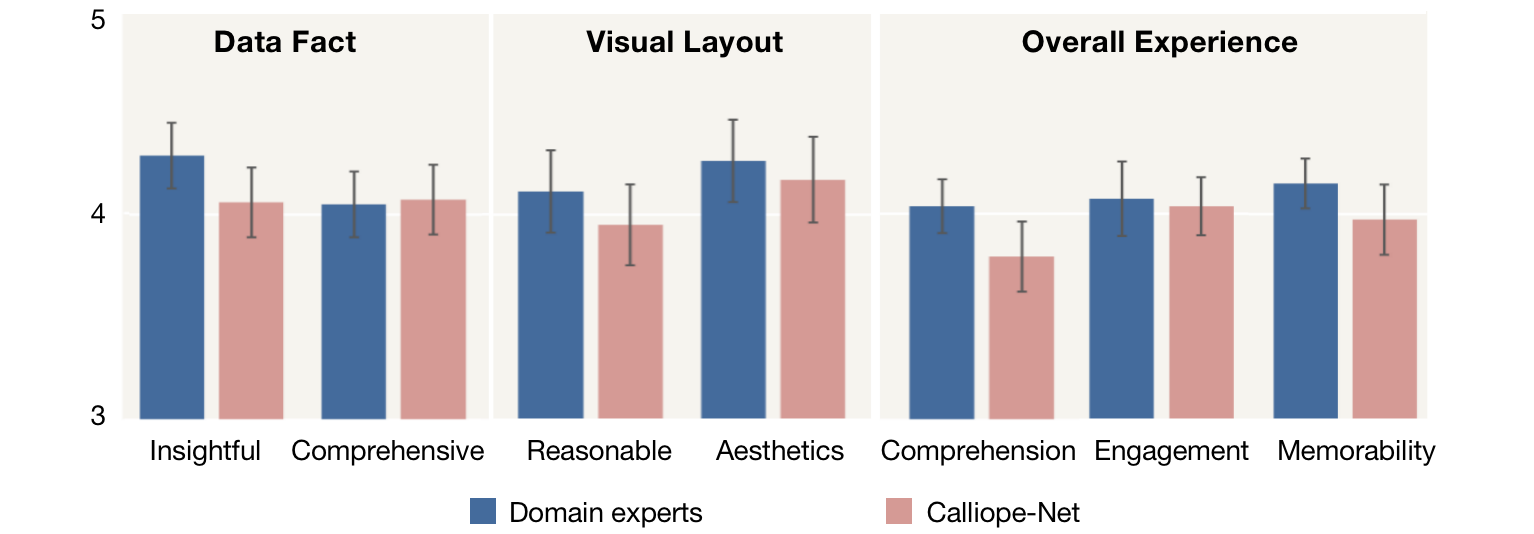}
    \vspace{-3mm}
    \caption{The ratings from the participants based on a 5-point Likert scale (1=strongly disagree, 5=strongly agree).}
    \label{fig:evaluation}
    \vspace{-6mm}
\end{figure}

\subsection{Feedback}
\rv{The participants assessed annotated graphs with mostly positive feedback, as shown in Fig.\ref{fig:evaluation}. 
Comparing the annotated graphs generated with those created by the experts, there is no significant difference ($p>.05$). The ratings for the generated graphs are almost as good as those created by the experts.} We also analyze subjective feedback to better understand the reasons for their ratings.

\textbf{Sufficient and Insightful Data Facts.} All the participants agreed that the extracted data facts were meaningful. 
P4 reported that 
\textit{``the information is rich as the graph covers many analysis perspectives''}. 
\new{P11 commented that \textit{``the main topic is clear and the facts are well organized with additional details .''}} V5 mentioned that \textit{``the annotated graph offers extensive information from a data perspective which can help quickly understand the data''}. V3 observed that \textit{``the annotated graph contains multiple levels of detail, covering the overall graph-level information as well as the community-level and the node-level insights''}. P8 also felt that \textit{``the number of annotations and the amount of information shown on the graph is adequate''}. \new{However, V12 mentioned that \textit{``the insights are relatively easy to understand but most of them are rather superficial''}. P10 noted that \textit{``since the specific meaning of community is not explained explicitly, I always wonder how the nodes are divided into different groups.''}}

\textbf{Expressive Visualizations and Effective Layout.} 
The participants appreciated the layout of the annotations, \textit{``the annotations make use of the blank space and do not overlap the important details in the graph''}(V2). P5 commented that \textit{``the distance from the annotation to the node is appropriate. I can easily locate the annotations with the corresponding graph elements''}. \new{P15 mentioned that \textit{``the type setting is very comfortable, and the annotations are placed in the middle so that the key points can be seen at a glance''}.} V5 noted that  \textit{``the annotations are placed from top to bottom. This placement is reasonable and fits normal reading habits''}. While most participants agreed the generated layout was elegant, we found that the annotated graphs created by experts would perform better in terms of the overall graph balance.

\textbf{Satisfactory Overall Experience.}
All the participants agreed that the annotated graph was clear and easy to understand. Compared the annotated graph generated by \name with that created by domain experts, we found experts were accustomed to creating narratives from overview to detail, while our algorithm focused more on the relationship between facts. Most participants noted that \textit{``the graph is well illustrated with detailed annotations''}(P1-P2, P5-P8, V1-V6, V9), \textit{``with the help of annotations, information about the node's degree, community's density are more explicit''}(V2). They also mentioned that the selection of related data facts and the organization improved general comprehension and memorability (P8, V3, V7, P13).

\subsection{Limitations and Future Work}
Apart from positive comments, we also found several limitations that can be improved in future work. 


\md{
\textbf{Facilitate customized designs and interactions.}
In future work, we plan to optimize our design goals to accommodate diverse and personalized needs. We plan to incorporate more interactive features into the system, such as adding options for matrix visualization, to enrich user experience and to provide greater flexibility.}

\md{\textbf{Enrich data semantics and text expressiveness.}
\name cannot extract the data semantics, such as the character relationships in a novel. Meanwhile, the text descriptions generated are still somewhat rigid, including some repetitive instances. 
The recent emergence of large language models (e.g., GPT-4) has provided possibilities for enhancing the readability and richness of the generated text descriptions.}

\textbf{Improve the scalability of \name.}
The maximum size of the input graph in \name is currently limited by the space of the poster and the efficiency of the algorithm. 
\rv{To better visualize larger graphs, we plan to incorporate advanced graph sampling and drawing methods~\cite{graph_sampling2,large_graph_vis,zhao2020preserving} 
into our current framework.}

\new{ \textbf{Integrate into the data story generation system.} Current system only supports node-link diagrams. Feedback from the user study indicated a need to combine tabular data with graph data. Thus, we plan to integrate the current prototype into the Calliope system to offer more comprehensive storytelling with both standard and graph visualizations.}

\section{Conclusion}
In this paper, we presented a framework and an interactive system for automatic fact discovery and presentation of graph data. \name employs a novel generation algorithm to gradually select graph data facts and organize them based on topic-evidence-explanation structure. An annotation-aware graph layout algorithm is applied to present facts with good readability. \md{Two} case studies and the user study demonstrate the capabilities of \name to extract graph data facts and present them with expressive annotated visualizations. We plan to further improve \name by addressing the limitations. 

\section*{Acknowledgments}
{This work was supported in part by the NSFC 62002267, 62072338, 62061136003, NSF Shanghai 23ZR1464700, and Shanghai Education Development Foundation ``Chen-Guang Project'' 21CGA75. This work was also supported by the Fundamental Research Funds for the Central Universities. We would like to thank Ziyan Liu for the interface design and our domain experts, Yier Shu, Manyun Zou, Yao Wei, Xingyu Lan, Yanhong Wu, and Qinghai Zhou for their valuable advice. We would also like to thank anonymous reviewers for their constructive feedback.}



\newpage

\bibliography{main}

\begin{thebibliography}{10}

\bibitem{ahn2013task}
J.-w. Ahn, C.~Plaisant, and B.~Shneiderman.
\newblock A task taxonomy for network evolution analysis.
\newblock {\em IEEE Transactions on Visualization and Computer Graphics},
  20(3):365--376, 2013. doi: {{%
10\hspace{.1pt}\discretionary{.}{%
}{.}\hspace{.4pt}1109\discretionary{/}{%
}{/}TVCG\hspace{.1pt}\discretionary{.}{%
}{.}\hspace{.4pt}2013\hspace{.1pt}\discretionary{.}{%
}{.}\hspace{.4pt}238}}


\bibitem{2016election}
G.~Aisch and K.~Yourish.
\newblock Connecting the dots behind the 2016 presidential candidates.
\newblock
  \url{https://www.nytimes.com/interactive/2015/05/17/us/elections/2016-presidential-campaigns-staff-connections-clinton-bush-cruz-paul-rubio-walker.html},
  June 2011.
\newblock Accessed: March 29, 2021.

\bibitem{bach2016telling}
B.~Bach, N.~Kerracher, K.~W. Hall, S.~Carpendale, J.~Kennedy, and
  N.~Henry~Riche.
\newblock Telling stories about dynamic networks with graph comics.
\newblock In {\em Proceedings of the ACM Conference on Human Factors in
  Computing Systems}, pp. 3670--3682, 2016. doi: {{%
10\hspace{.1pt}\discretionary{.}{%
}{.}\hspace{.4pt}1145\discretionary{/}{%
}{/}2858036\hspace{.1pt}\discretionary{.}{%
}{.}\hspace{.4pt}2858387}}


\bibitem{bastian2009gephi}
M.~Bastian, S.~Heymann, and M.~Jacomy.
\newblock Gephi: An open source software for exploring and manipulating
  networks.
\newblock {\em Proceedings of the International AAAI Conference on Web and
  Social Media}, 3(1), Mar. 2009. doi: {{%
10\hspace{.1pt}\discretionary{.}{%
}{.}\hspace{.4pt}13140\discretionary{/}{%
}{/}2\hspace{.1pt}\discretionary{.}{%
}{.}\hspace{.4pt}1\hspace{.1pt}\discretionary{.}{%
}{.}\hspace{.4pt}1341\hspace{.1pt}\discretionary{.}{%
}{.}\hspace{.4pt}1520}}


\bibitem{NLG_visualization}
S.~L.~F. Beck.
\newblock Vis author profiles: Interactive descriptions of publication records
  combining text and visualization.
\newblock {\em IEEE Transactions on Visualization and Computer Graphics},
  25(1):152--161, 2018. doi: {{%
10\hspace{.1pt}\discretionary{.}{%
}{.}\hspace{.4pt}1109\discretionary{/}{%
}{/}TVCG\hspace{.1pt}\discretionary{.}{%
}{.}\hspace{.4pt}2018\hspace{.1pt}\discretionary{.}{%
}{.}\hspace{.4pt}2865022}}


\bibitem{bennett2007aesthetics}
C.~Bennett, J.~Ryall, L.~Spalteholz, and A.~A. Gooch.
\newblock The aesthetics of graph visualization.
\newblock In {\em Proceedings of the Computational Aesthetics in Graphics,
  Visualization, and Imaging}, pp. 57--64, 2007. doi: {{%
10\hspace{.1pt}\discretionary{.}{%
}{.}\hspace{.4pt}2312\discretionary{/}{%
}{/}compaesth\discretionary{/}{%
}{/}compaesth07\discretionary{/}{%
}{/}057\discretionary{%
}{-}{-}064}}


\bibitem{bonacich1987power}
P.~Bonacich.
\newblock Power and centrality: A family of measures.
\newblock {\em American journal of sociology}, 92(5):1170--1182, 1987. doi: {{%
10\hspace{.1pt}\discretionary{.}{%
}{.}\hspace{.4pt}1086\discretionary{/}{%
}{/}228631}}


\bibitem{bounegru2017narrating}
L.~Bounegru, T.~Venturini, J.~Gray, and M.~Jacomy.
\newblock Narrating networks: Exploring the affordances of networks as
  storytelling devices in journalism.
\newblock {\em Digital Journalism}, 5(6):699--730, 2017. doi: {{%
10\hspace{.1pt}\discretionary{.}{%
}{.}\hspace{.4pt}1080\discretionary{/}{%
}{/}21670811\hspace{.1pt}\discretionary{.}{%
}{.}\hspace{.4pt}2016\hspace{.1pt}\discretionary{.}{%
}{.}\hspace{.4pt}1186497}}


\bibitem{brath2018automated}
R.~Brath and M.~Matusiak.
\newblock Automated annotations.
\newblock In {\em An IEEE VIS workshop on visualization for communication
  (VisComm)}, 2018.

\bibitem{bryan2016temporal}
C.~Bryan, K.-L. Ma, and J.~Woodring.
\newblock Temporal summary images: An approach to narrative visualization via
  interactive annotation generation and placement.
\newblock {\em IEEE Transactions on Visualization and Computer Graphics},
  23(1):511--520, 2016. doi: {{%
10\hspace{.1pt}\discretionary{.}{%
}{.}\hspace{.4pt}1109\discretionary{/}{%
}{/}TVCG\hspace{.1pt}\discretionary{.}{%
}{.}\hspace{.4pt}2016\hspace{.1pt}\discretionary{.}{%
}{.}\hspace{.4pt}2598876}}


\bibitem{campbell2017detailed}
E.~M. Campbell, H.~Jia, A.~Shankar, D.~Hanson, W.~Luo, S.~Masciotra, S.~M.
  Owen, A.~M. Oster, R.~R. Galang, M.~W. Spiller, et~al.
\newblock Detailed transmission network analysis of a large opiate-driven
  outbreak of hiv infection in the united states.
\newblock {\em The Journal of Infectious Diseases}, 216(9):1053--1062, 2017.
  doi: {{%
10\hspace{.1pt}\discretionary{.}{%
}{.}\hspace{.4pt}1093\discretionary{/}{%
}{/}infdis\discretionary{/}{%
}{/}jix307}}


\bibitem{chabot2003tableau}
C.~Chabot, C.~Stolte, and P.~Hanrahan.
\newblock Tableau software.
\newblock {\em Tableau Software}, 6, 2003.

\bibitem{chen2023does}
Q.~Chen, S.~Cao, J.~Wang, and N.~Cao.
\newblock How does automation shape the process of narrative visualization: A
  survey of tools.
\newblock {\em IEEE Transactions on Visualization and Computer Graphics}, 2023.
  doi: {{%
10\hspace{.1pt}\discretionary{.}{%
}{.}\hspace{.4pt}1109\discretionary{/}{%
}{/}TVCG\hspace{.1pt}\discretionary{.}{%
}{.}\hspace{.4pt}2023\hspace{.1pt}\discretionary{.}{%
}{.}\hspace{.4pt}3261320}}


\bibitem{Automated_Infographic}
Z.~Chen, Y.~Wang, Q.~Wang, Y.~Wang, and H.~Qu.
\newblock Towards automated infographic design: Deep learning-based
  auto-extraction of extensible timeline.
\newblock {\em IEEE Transactions on Visualization and Computer Graphics},
  26(1):917--926, 2020. doi: {{%
10\hspace{.1pt}\discretionary{.}{%
}{.}\hspace{.4pt}1109\discretionary{/}{%
}{/}TVCG\hspace{.1pt}\discretionary{.}{%
}{.}\hspace{.4pt}2019\hspace{.1pt}\discretionary{.}{%
}{.}\hspace{.4pt}2934810}}


\bibitem{choe2015characterizing}
E.~K. Choe, B.~Lee, et~al.
\newblock Characterizing visualization insights from quantified selfers'
  personal data presentations.
\newblock {\em IEEE computer graphics and applications}, 35(4):28--37, 2015.
  doi: {{%
10\hspace{.1pt}\discretionary{.}{%
}{.}\hspace{.4pt}1109\discretionary{/}{%
}{/}MCG\hspace{.1pt}\discretionary{.}{%
}{.}\hspace{.4pt}2015\hspace{.1pt}\discretionary{.}{%
}{.}\hspace{.4pt}51}}


\bibitem{clauset2004finding}
A.~Clauset, M.~E. Newman, and C.~Moore.
\newblock Finding community structure in very large networks.
\newblock {\em Phys. Rev. E}, 70:066111, Dec 2004. doi: {{%
10\hspace{.1pt}\discretionary{.}{%
}{.}\hspace{.4pt}1103\discretionary{/}{%
}{/}PhysRevE\hspace{.1pt}\discretionary{.}{%
}{.}\hspace{.4pt}70\hspace{.1pt}\discretionary{.}{%
}{.}\hspace{.4pt}066111}}


\bibitem{bushmoney}
S.~Cohen, L.~Spirito, and A.~Crites.
\newblock The bush money machine.
\newblock
  \url{http://www.washingtonpost.com/wp-srv/politics/pioneers/network_graphic.pdf},
  May 2004.
\newblock Accessed: March 29, 2021.

\bibitem{demiralp2017foresight}
{\c{C}}.~Demiralp, P.~J. Haas, S.~Parthasarathy, and T.~Pedapati.
\newblock Foresight: Recommending visual insights.
\newblock {\em arXiv preprint arXiv:1707.03877}, 2017.

\bibitem{deodhar2022humanml}
M.~Deodhar, X.~Ma, Y.~Cai, A.~Koes, A.~Beutel, and J.~Chen.
\newblock A human-ml collaboration framework for improving video content
  reviews.
\newblock {\em arXiv preprint arXiv:2210.09500}, 2022.

\bibitem{ding2019quickinsights}
R.~Ding, S.~Han, Y.~Xu, H.~Zhang, and D.~Zhang.
\newblock Quickinsights: Quick and automatic discovery of insights from
  multi-dimensional data.
\newblock In {\em Proceedings of the ACM International Conference on Management
  of Data}, pp. 317--332, Jun 2019. doi: {{%
10\hspace{.1pt}\discretionary{.}{%
}{.}\hspace{.4pt}1145\discretionary{/}{%
}{/}3299869\hspace{.1pt}\discretionary{.}{%
}{.}\hspace{.4pt}3314037}}


\bibitem{eiter1994computing}
T.~Eiter and H.~Mannila.
\newblock Computing discrete fr{\'e}chet distance.
\newblock Technical report, Citeseer, 1994.

\bibitem{ellson2001graphviz}
J.~Ellson, E.~Gansner, L.~Koutsofios, S.~C. North, and G.~Woodhull.
\newblock Graphviz—open source graph drawing tools.
\newblock In {\em International Symposium on Graph Drawing}, pp. 483--484.
  Springer, Feb 2001. doi: {{%
10\hspace{.1pt}\discretionary{.}{%
}{.}\hspace{.4pt}1007\discretionary{/}{%
}{/}3\discretionary{%
}{-}{-}540\discretionary{%
}{-}{-}45848\discretionary{%
}{-}{-}4\_57}}


\bibitem{flake2000efficient}
G.~W. Flake, S.~Lawrence, and C.~L. Giles.
\newblock Efficient identification of web communities.
\newblock In {\em Proceedings of the ACM SIGKDD International Conference on
  Knowledge Discovery and Data Mining}, pp. 150--160, Aug 2000. doi: {{%
10\hspace{.1pt}\discretionary{.}{%
}{.}\hspace{.4pt}1145\discretionary{/}{%
}{/}347090\hspace{.1pt}\discretionary{.}{%
}{.}\hspace{.4pt}347121}}


\bibitem{fortunato2010community}
S.~Fortunato.
\newblock Community detection in graphs.
\newblock {\em Physics Reports}, 486(3-5):75--174, 2010. doi: {{%
10\hspace{.1pt}\discretionary{.}{%
}{.}\hspace{.4pt}1016\discretionary{/}{%
}{/}j\hspace{.1pt}\discretionary{.}{%
}{.}\hspace{.4pt}physrep\hspace{.1pt}\discretionary{.}{%
}{.}\hspace{.4pt}2009\hspace{.1pt}\discretionary{.}{%
}{.}\hspace{.4pt}11\hspace{.1pt}\discretionary{.}{%
}{.}\hspace{.4pt}002}}


\bibitem{freeman2004development}
L.~Freeman.
\newblock The development of social network analysis.
\newblock {\em A Study in the Sociology of Science}, 1(687):159--167, Jan 2004.

\bibitem{gansner2004graph}
E.~R. Gansner, Y.~Koren, and S.~North.
\newblock Graph drawing by stress majorization.
\newblock In {\em International Symposium on Graph Drawing}, pp. 239--250.
  Springer, 2004.

\bibitem{geem2001new}
Z.~W. Geem, J.~H. Kim, and G.~V. Loganathan.
\newblock A new heuristic optimization algorithm: Harmony search.
\newblock {\em Simulation}, 76(2):60--68, 2001. doi: {{%
10\hspace{.1pt}\discretionary{.}{%
}{.}\hspace{.4pt}1177\discretionary{/}{%
}{/}003754970107600201}}


\bibitem{gower1975generalized}
J.~C. Gower.
\newblock Generalized procrustes analysis.
\newblock {\em Psychometrika}, 40(1):33--51, 1975.

\bibitem{hansen2010analyzing}
D.~Hansen, B.~Shneiderman, and M.~A. Smith.
\newblock Analyzing social media networks with {NodeXL}: Insights from a
  connected world.
\newblock {\em International Journal of Human-Computer Interaction},
  27(4):405--408, Feb 2010. doi: {{%
10447318\hspace{.1pt}\discretionary{.}{%
}{.}\hspace{.4pt}2011\hspace{.1pt}\discretionary{.}{%
}{.}\hspace{.4pt}544971}}


\bibitem{hart2021storycraft}
J.~Hart.
\newblock {\em Storycraft: The complete guide to writing narrative nonfiction}.
\newblock University of Chicago Press, 2021.

\bibitem{herman2000graph}
I.~Herman, G.~Melan{\c{c}}on, and M.~S. Marshall.
\newblock Graph visualization and navigation in information visualization: A
  survey.
\newblock {\em IEEE Transactions on Visualization and Computer Graphics},
  6(1):24--43, 2000. doi: {{%
10\hspace{.1pt}\discretionary{.}{%
}{.}\hspace{.4pt}1109\discretionary{/}{%
}{/}2945\hspace{.1pt}\discretionary{.}{%
}{.}\hspace{.4pt}841119}}


\bibitem{huang2005layout}
W.~Huang, P.~Eades, and S.-H. Hong.
\newblock {\em Layout effects: Comparison of sociogram drawing conventions}.
\newblock School of Information Technologies, University of Sydney Darlington,
  England, Jan 2005.

\bibitem{hullman2013contextifier}
J.~Hullman, N.~Diakopoulos, and E.~Adar.
\newblock Contextifier: automatic generation of annotated stock visualizations.
\newblock In {\em Proceedings of the ACM Conference on Human Factors in
  Computing Systems}, pp. 2707--2716, 2013. doi: {{%
10\hspace{.1pt}\discretionary{.}{%
}{.}\hspace{.4pt}1145\discretionary{/}{%
}{/}2470654\hspace{.1pt}\discretionary{.}{%
}{.}\hspace{.4pt}2481374}}


\bibitem{hullman2013deeper}
J.~Hullman, S.~Drucker, N.~H. Riche, B.~Lee, D.~Fisher, and E.~Adar.
\newblock A deeper understanding of sequence in narrative visualization.
\newblock {\em IEEE Transactions on Visualization and Computer Graphics},
  19(12):2406--2415, 2013. doi: {{%
10\hspace{.1pt}\discretionary{.}{%
}{.}\hspace{.4pt}1109\discretionary{/}{%
}{/}TVCG\hspace{.1pt}\discretionary{.}{%
}{.}\hspace{.4pt}2013\hspace{.1pt}\discretionary{.}{%
}{.}\hspace{.4pt}119}}


\bibitem{socialpark}
J.~Ilbo.
\newblock Social network analysis of high-ranking officials in s. korean
  government.
\newblock \url{https://www.ire.org/product/story-25691/}, 2012.
\newblock Accessed: March 29, 2021.

\bibitem{kaneider2013automatic}
D.~Kaneider, T.~Seifried, and M.~Haller.
\newblock Automatic annotation placement for interactive maps.
\newblock In {\em Proceedings of the ACM International Conference on
  Interactive Tabletops and Surfaces}, pp. 61--70. Association for Computing
  Machinery, 2013. doi: {{%
10\hspace{.1pt}\discretionary{.}{%
}{.}\hspace{.4pt}1145\discretionary{/}{%
}{/}2512349\hspace{.1pt}\discretionary{.}{%
}{.}\hspace{.4pt}2512809}}


\bibitem{kim2019datatoon}
N.~W. Kim, N.~Henry~Riche, B.~Bach, G.~Xu, M.~Brehmer, K.~Hinckley, M.~Pahud,
  H.~Xia, M.~J. McGuffin, and H.~Pfister.
\newblock Datatoon: Drawing dynamic network comics with pen+touch interaction.
\newblock In {\em Proceedings of the ACM Conference on Human Factors in
  Computing Systems}, pp. 1--12, 2019. doi: {{%
10\hspace{.1pt}\discretionary{.}{%
}{.}\hspace{.4pt}1145\discretionary{/}{%
}{/}3290605\hspace{.1pt}\discretionary{.}{%
}{.}\hspace{.4pt}3300335}}


\bibitem{knaflic2015storytelling}
C.~N. Knaflic.
\newblock {\em Storytelling with data: A data visualization guide for business
  professionals}.
\newblock John Wiley \& Sons, 2015.

\bibitem{knuth1993stanford}
D.~E. Knuth.
\newblock The stanford graphbase: a platform for combinatorial algorithms.
\newblock In {\em SODA}, vol.~93, pp. 41--43, 1993.

\bibitem{kosara2013storytelling}
R.~Kosara and J.~Mackinlay.
\newblock Storytelling: The next step for visualization.
\newblock {\em Computer}, 46(5):44--50, 2013. doi: {{%
10\hspace{.1pt}\discretionary{.}{%
}{.}\hspace{.4pt}1109\discretionary{/}{%
}{/}MC\hspace{.1pt}\discretionary{.}{%
}{.}\hspace{.4pt}2013\hspace{.1pt}\discretionary{.}{%
}{.}\hspace{.4pt}36}}


\bibitem{latif2019EuroVis}
S.~Latif, K.~Su, and F.~Beck.
\newblock Authoring combined textual and visual descriptions of graph data.
\newblock In {\em EuroVis (Short Papers)}, pp. 115--119, 2019.

\bibitem{latif2021kori}
S.~Latif, Z.~Zhou, Y.~Kim, F.~Beck, and N.~W. Kim.
\newblock Kori: Interactive synthesis of text and charts in data documents.
\newblock {\em IEEE Transactions on Visualization and Computer Graphics},
  28(1):184--194, 2021. doi: {{%
10\hspace{.1pt}\discretionary{.}{%
}{.}\hspace{.4pt}1109\discretionary{/}{%
}{/}TVCG\hspace{.1pt}\discretionary{.}{%
}{.}\hspace{.4pt}2021\hspace{.1pt}\discretionary{.}{%
}{.}\hspace{.4pt}3114802}}


\bibitem{law2020characterizing}
P.-M. Law, A.~Endert, and J.~Stasko.
\newblock Characterizing automated data insights.
\newblock In {\em 2020 IEEE Visualization Conference (VIS)}, pp. 171--175.
  IEEE, 2020. doi: {{%
10\hspace{.1pt}\discretionary{.}{%
}{.}\hspace{.4pt}1109\discretionary{/}{%
}{/}VIS47514\hspace{.1pt}\discretionary{.}{%
}{.}\hspace{.4pt}2020\hspace{.1pt}\discretionary{.}{%
}{.}\hspace{.4pt}00041}}


\bibitem{lazar2017research}
J.~Lazar, J.~H. Feng, and H.~Hochheiser.
\newblock {\em Research methods in human-computer interaction}.
\newblock Morgan Kaufmann, 2017.

\bibitem{lee_task_2006}
B.~Lee, C.~Plaisant, C.~S. Parr, J.-D. Fekete, and N.~Henry.
\newblock Task taxonomy for graph visualization.
\newblock In {\em Proceedings of the 2006 {AVI} workshop on {BEyond} Time and
  Errors Novel Evaluation Methods for Information Visualization - {BELIV} '06},
  p.~1. {ACM} Press, 2006. doi: {{%
10\hspace{.1pt}\discretionary{.}{%
}{.}\hspace{.4pt}1145\discretionary{/}{%
}{/}1168149\hspace{.1pt}\discretionary{.}{%
}{.}\hspace{.4pt}1168168}}


\bibitem{lee2015more}
B.~Lee, N.~H. Riche, P.~Isenberg, and S.~Carpendale.
\newblock More than telling a story: Transforming data into visually shared
  stories.
\newblock {\em IEEE computer graphics and applications}, 35(5):84--90, 2015.
  doi: {{%
10\hspace{.1pt}\discretionary{.}{%
}{.}\hspace{.4pt}1109\discretionary{/}{%
}{/}MCG\hspace{.1pt}\discretionary{.}{%
}{.}\hspace{.4pt}2015\hspace{.1pt}\discretionary{.}{%
}{.}\hspace{.4pt}99}}


\bibitem{lee2019avoiding}
D.~J.-L. Lee, H.~Dev, H.~Hu, H.~Elmeleegy, and A.~Parameswaran.
\newblock Avoiding drill-down fallacies with {VisPilot}: Assisted exploration
  of data subsets.
\newblock In {\em Proceedings of the 24th International Conference on
  Intelligent User Interfaces}, pp. 186--196, march 2019. doi: {{%
10\hspace{.1pt}\discretionary{.}{%
}{.}\hspace{.4pt}1145\discretionary{/}{%
}{/}3301275\hspace{.1pt}\discretionary{.}{%
}{.}\hspace{.4pt}3302307}}


\bibitem{lu2020exploring}
M.~Lu, C.~Wang, J.~Lanir, N.~Zhao, H.~Pfister, D.~Cohen-Or, and H.~Huang.
\newblock Exploring visual information flows in infographics.
\newblock In {\em Proceedings of the ACM Conference on Human Factors in
  Computing Systems}, pp. 1--12, 2020. doi: {{%
10\hspace{.1pt}\discretionary{.}{%
}{.}\hspace{.4pt}1145\discretionary{/}{%
}{/}3313831\hspace{.1pt}\discretionary{.}{%
}{.}\hspace{.4pt}3376263}}


\bibitem{luo2020ICDE}
Y.~Luo, C.~Chai, X.~Qin, N.~Tang, and G.~Li.
\newblock Interactive cleaning for progressive visualization through composite
  questions.
\newblock In {\em 2020 IEEE 36th International Conference on Data Engineering
  (ICDE)}, pp. 733--744. IEEE, 2020. doi: {{%
10\hspace{.1pt}\discretionary{.}{%
}{.}\hspace{.4pt}1109\discretionary{/}{%
}{/}ICDE48307\hspace{.1pt}\discretionary{.}{%
}{.}\hspace{.4pt}2020\hspace{.1pt}\discretionary{.}{%
}{.}\hspace{.4pt}00069}}


\bibitem{mafrur2018ACM}
R.~Mafrur, M.~A. Sharaf, and H.~A. Khan.
\newblock Dive: Diversifying view recommendation for visual data exploration.
\newblock In {\em Proceedings of the 27th ACM International Conference on
  Information and Knowledge Management}, pp. 1123--1132, 2018. doi: {{%
10\hspace{.1pt}\discretionary{.}{%
}{.}\hspace{.4pt}1145\discretionary{/}{%
}{/}3269206\hspace{.1pt}\discretionary{.}{%
}{.}\hspace{.4pt}3271744}}


\bibitem{chris}
B.~Marsh and K.~Zernike.
\newblock Chris christie and the lane closings: A spectator’s guide.
\newblock
  \url{https://www.nytimes.com/interactive/2015/04/08/nyregion/chris-christie-and-bridgegate-guide.html},
  Apr. 2015.
\newblock Accessed: March 29, 2021.

\bibitem{nettleton2013data}
D.~F. Nettleton.
\newblock Data mining of social networks represented as graphs.
\newblock {\em Computer Science Review}, 7:1--34, 2013. doi: {{%
10\hspace{.1pt}\discretionary{.}{%
}{.}\hspace{.4pt}1016\discretionary{/}{%
}{/}j\hspace{.1pt}\discretionary{.}{%
}{.}\hspace{.4pt}cosrev\hspace{.1pt}\discretionary{.}{%
}{.}\hspace{.4pt}2012\hspace{.1pt}\discretionary{.}{%
}{.}\hspace{.4pt}12\hspace{.1pt}\discretionary{.}{%
}{.}\hspace{.4pt}001}}


\bibitem{nobre2019state}
C.~Nobre, M.~Meyer, M.~Streit, and A.~Lex.
\newblock The state of the art in visualizing multivariate networks.
\newblock {\em Computer Graphics Forum}, 38(3):807--832, 2019.

\bibitem{proxy}
OCCRP.
\newblock The proxy platform.
\newblock \url{https://www.reportingproject.net/proxy/en/}, 2011.
\newblock Accessed: March 29, 2021.

\bibitem{large_graph_vis}
K.-L.~M. Oh-Hyun~Kwon, Tarik~Crnovrsanin.
\newblock What would a graph look like in this layout? a machine learning
  approach to large graph visualization.
\newblock {\em IEEE Transactions on Visualization and Computer Graphics},
  24:478--488, 2017. doi: {{%
10\hspace{.1pt}\discretionary{.}{%
}{.}\hspace{.4pt}1109\discretionary{/}{%
}{/}TVCG\hspace{.1pt}\discretionary{.}{%
}{.}\hspace{.4pt}2017\hspace{.1pt}\discretionary{.}{%
}{.}\hspace{.4pt}2743858}}


\bibitem{pohl2009comparing}
M.~Pohl, M.~Schmitt, and S.~Diehl.
\newblock Comparing the readability of graph layouts using eyetracking and
  task-oriented analysis.
\newblock In {\em Computational Aesthetics in Graphics, Visualization, and
  Imaging}, pp. 49--56, 2009. doi: {{%
10\hspace{.1pt}\discretionary{.}{%
}{.}\hspace{.4pt}2312\discretionary{/}{%
}{/}COMPAESTH\discretionary{/}{%
}{/}COMPAESTH09\discretionary{/}{%
}{/}049\discretionary{%
}{-}{-}056}}


\bibitem{pretorius2014tasks}
J.~Pretorius, H.~C. Purchase, and J.~T. Stasko.
\newblock Tasks for multivariate network analysis.
\newblock In {\em Multivariate Network Visualization}, pp. 77--95. Springer,
  2014.

\bibitem{purchase2000effective}
H.~C. Purchase.
\newblock Effective information visualisation: a study of graph drawing
  aesthetics and algorithms.
\newblock {\em Interacting with Computers}, 13(2):147--162, 2000. doi: {{%
10\hspace{.1pt}\discretionary{.}{%
}{.}\hspace{.4pt}1016\discretionary{/}{%
}{/}S0953\discretionary{%
}{-}{-}5438\discretionary{%
}{(}{(}00\discretionary{)}{%
}{)}00032\discretionary{%
}{-}{-}1}}


\bibitem{ren2017chartaccent}
D.~Ren, M.~Brehmer, B.~Lee, T.~H{\"o}llerer, and E.~K. Choe.
\newblock Chartaccent: Annotation for data-driven storytelling.
\newblock In {\em IEEE Pacific Visualization Symposium (PacificVis)}, pp.
  230--239. IEEE, 2017. doi: {{%
10\hspace{.1pt}\discretionary{.}{%
}{.}\hspace{.4pt}1109\discretionary{/}{%
}{/}PACIFICVIS\hspace{.1pt}\discretionary{.}{%
}{.}\hspace{.4pt}2017\hspace{.1pt}\discretionary{.}{%
}{.}\hspace{.4pt}8031599}}


\bibitem{romat2019expressive}
H.~Romat, C.~Appert, and E.~Pietriga.
\newblock Expressive authoring of node-link diagrams with graphies.
\newblock {\em IEEE Transactions on Visualization and Computer Graphics},
  27(4):2329--2340, 2019. doi: {{%
10\hspace{.1pt}\discretionary{.}{%
}{.}\hspace{.4pt}1109\discretionary{/}{%
}{/}TVCG\hspace{.1pt}\discretionary{.}{%
}{.}\hspace{.4pt}2019\hspace{.1pt}\discretionary{.}{%
}{.}\hspace{.4pt}2950932}}


\bibitem{segel2010narrative}
E.~Segel and J.~Heer.
\newblock Narrative visualization: Telling stories with data.
\newblock {\em IEEE Transactions on Visualization and Computer Graphics},
  16(6):1139--1148, 2010. doi: {{%
10\hspace{.1pt}\discretionary{.}{%
}{.}\hspace{.4pt}1109\discretionary{/}{%
}{/}TVCG\hspace{.1pt}\discretionary{.}{%
}{.}\hspace{.4pt}2010\hspace{.1pt}\discretionary{.}{%
}{.}\hspace{.4pt}179}}


\bibitem{shannon2003cytoscape}
P.~Shannon, A.~Markiel, O.~Ozier, N.~S. Baliga, J.~T. Wang, D.~Ramage, N.~Amin,
  B.~Schwikowski, and T.~Ideker.
\newblock Cytoscape: a software environment for integrated models of
  biomolecular interaction networks.
\newblock {\em Genome Research}, 13(11):2498--2504, 2003. doi: {{%
10\hspace{.1pt}\discretionary{.}{%
}{.}\hspace{.4pt}1101\discretionary{/}{%
}{/}gr\hspace{.1pt}\discretionary{.}{%
}{.}\hspace{.4pt}1239303}}


\bibitem{shi2020calliope}
D.~Shi, X.~Xu, F.~Sun, Y.~Shi, and N.~Cao.
\newblock Calliope: Automatic visual data story generation from a spreadsheet.
\newblock {\em IEEE Transactions on Visualization and Computer Graphics},
  27(2):453--463, 2021. doi: {{%
10\hspace{.1pt}\discretionary{.}{%
}{.}\hspace{.4pt}1109\discretionary{/}{%
}{/}TVCG\hspace{.1pt}\discretionary{.}{%
}{.}\hspace{.4pt}2020\hspace{.1pt}\discretionary{.}{%
}{.}\hspace{.4pt}3030403}}


\bibitem{shi2000normalized}
J.~Shi and J.~Malik.
\newblock Normalized cuts and image segmentation.
\newblock {\em IEEE Transactions on Pattern Analysis and Machine Intelligence},
  22(8):888--905, 2000. doi: {{%
10\hspace{.1pt}\discretionary{.}{%
}{.}\hspace{.4pt}1109\discretionary{/}{%
}{/}34\hspace{.1pt}\discretionary{.}{%
}{.}\hspace{.4pt}868688}}


\bibitem{srinivasan_augmenting_2019}
A.~Srinivasan, S.~M. Drucker, A.~Endert, and J.~Stasko.
\newblock Augmenting visualizations with interactive data facts to facilitate
  interpretation and communication.
\newblock {\em IEEE Transactions on Visualization and Computer Graphics},
  25(1):672--681, 2019. doi: {{%
10\hspace{.1pt}\discretionary{.}{%
}{.}\hspace{.4pt}1109\discretionary{/}{%
}{/}TVCG\hspace{.1pt}\discretionary{.}{%
}{.}\hspace{.4pt}2018\hspace{.1pt}\discretionary{.}{%
}{.}\hspace{.4pt}2865145}}


\bibitem{stackoverflow}
{Stack Overflow}.
\newblock Stack overflow tag network.
\newblock
  \url{https://www.kaggle.com/stackoverflow/stack-overflow-tag-network}, 2021.
\newblock Accessed: March 15, 2021.

\bibitem{tang2017extracting}
B.~Tang, S.~Han, M.~L. Yiu, R.~Ding, and D.~Zhang.
\newblock Extracting top-k insights from multi-dimensional data.
\newblock In {\em Proceedings of the ACM International Conference on Management
  of Data}, pp. 1509--1524, 2017. doi: {{%
10\hspace{.1pt}\discretionary{.}{%
}{.}\hspace{.4pt}1145\discretionary{/}{%
}{/}3035918\hspace{.1pt}\discretionary{.}{%
}{.}\hspace{.4pt}3035922}}


\bibitem{london-riots}
G.~I. team, R.~Procter, F.~Vis, and A.~Voss.
\newblock How riot rumours spread on twitter.
\newblock
  \url{https://www.politico.eu/interactive/lobbyists-brussels-social-network-meetings-commission-strategy/},
  Dec. 2011.
\newblock Accessed: March 29, 2021.

\bibitem{vartak_seedb_2015}
M.~Vartak, S.~Rahman, S.~Madden, A.~Parameswaran, and N.~Polyzotis.
\newblock Seedb: Efficient data-driven visualization recommendations to support
  visual analytics.
\newblock In {\em Proceedings of the VLDB Endowment International Conference on
  Very Large Data Bases}, vol.~8, p. 2182. NIH Public Access, 2015.

\bibitem{ttip}
J.~Von~Daniels, O.~Marta, M.~Klack, S.~Wehrmeyer, and S.~Jockers.
\newblock Die ttip-dealer.
\newblock \url{https://correctiv.org/recherchen/ttip/dealer/}.
\newblock Accessed: March 29, 2021.

\bibitem{wang2019datashot}
Y.~Wang, Z.~Sun, H.~Zhang, W.~Cui, K.~Xu, X.~Ma, and D.~Zhang.
\newblock {DataShot}: Automatic generation of fact sheets from tabular data.
\newblock {\em IEEE Transactions on Visualization and Computer Graphics},
  26(1):895--905, 2019. doi: {{%
10\hspace{.1pt}\discretionary{.}{%
}{.}\hspace{.4pt}1109\discretionary{/}{%
}{/}TVCG\hspace{.1pt}\discretionary{.}{%
}{.}\hspace{.4pt}2019\hspace{.1pt}\discretionary{.}{%
}{.}\hspace{.4pt}2934398}}


\bibitem{wang2017revisiting}
Y.~Wang, Y.~Wang, Y.~Sun, L.~Zhu, K.~Lu, C.-W. Fu, M.~Sedlmair, O.~Deussen, and
  B.~Chen.
\newblock Revisiting stress majorization as a unified framework for interactive
  constrained graph visualization.
\newblock {\em IEEE Transactions on Visualization and Computer Graphics},
  24(1):489--499, 2017. doi: {{%
10\hspace{.1pt}\discretionary{.}{%
}{.}\hspace{.4pt}1109\discretionary{/}{%
}{/}TVCG\hspace{.1pt}\discretionary{.}{%
}{.}\hspace{.4pt}2017\hspace{.1pt}\discretionary{.}{%
}{.}\hspace{.4pt}2745919}}


\bibitem{watts1998collective}
D.~J. Watts and S.~H. Strogatz.
\newblock Collective dynamics of ‘small-world’ networks.
\newblock {\em nature}, 393(6684):440--442, 1998.

\bibitem{enwiki:1006098025}
{Wikipedia contributors}.
\newblock Les misérables --- {Wikipedia}{,} the free encyclopedia.
\newblock
  \url{https://en.wikipedia.org/w/index.php?title=Les_Mis%C3%A9rables&oldid=1006098025},
  2021.
\newblock Accessed: March 31, 2021.

\bibitem{wu2016focus+}
H.-Y. Wu.
\newblock Focus+context metro map layout and annotation.
\newblock In {\em Proceedings of the 32nd Spring Conference on Computer
  Graphics}, pp. 41--47, 2016.

\bibitem{yang_chen_toward_2009}
{Yang Chen}, {Jing Yang}, and W.~Ribarsky.
\newblock Toward effective insight management in visual analytics systems.
\newblock In {\em {IEEE} Pacific Visualization Symposium (PacificVis)}, pp.
  49--56, Apr. 2009.
\newblock ISSN: 2165-8773. doi: {{%
10\hspace{.1pt}\discretionary{.}{%
}{.}\hspace{.4pt}1109\discretionary{/}{%
}{/}PACIFICVIS\hspace{.1pt}\discretionary{.}{%
}{.}\hspace{.4pt}2009\hspace{.1pt}\discretionary{.}{%
}{.}\hspace{.4pt}4906837}}


\bibitem{graph_sampling2}
B.~Z. Zhang~Peng.
\newblock A classification sampling algorithm over the dynamic streaming graph.
\newblock {\em 2022 4th International Conference on Advances in Computer
  Technology, Information Science and Communications (CTISC)}, 2022. doi: {{%
10\hspace{.1pt}\discretionary{.}{%
}{.}\hspace{.4pt}1109\discretionary{/}{%
}{/}CTISC54888\hspace{.1pt}\discretionary{.}{%
}{.}\hspace{.4pt}2022\hspace{.1pt}\discretionary{.}{%
}{.}\hspace{.4pt}9849754}}


\bibitem{zhao2020preserving}
Y.~Zhao, H.~Jiang, Y.~Qin, H.~Xie, Y.~Wu, S.~Liu, Z.~Zhou, J.~Xia, F.~Zhou,
  et~al.
\newblock Preserving minority structures in graph sampling.
\newblock {\em IEEE Transactions on Visualization and Computer Graphics},
  27(2):1698--1708, 2020. doi: {{%
10\hspace{.1pt}\discretionary{.}{%
}{.}\hspace{.4pt}1109\discretionary{/}{%
}{/}TVCG\hspace{.1pt}\discretionary{.}{%
}{.}\hspace{.4pt}2020\hspace{.1pt}\discretionary{.}{%
}{.}\hspace{.4pt}3030428}}


\end{thebibliography}
\bibliographystyle{abbrv-doi}

\clearpage

\end{document}